\def\isabridged{1}   

\documentclass[conference]{IEEEtran}
\usepackage{caption}
\usepackage{subcaption}
\usepackage{listings}
\usepackage{paralist}
\usepackage{enumitem}
\usepackage{pifont}
\usepackage{xspace}
\usepackage{comment}
\usepackage{fancyvrb}
\usepackage{graphicx,txfonts}
\usepackage{multirow}
\usepackage{float}
\usepackage{tikz}
\usepackage{multicol}
\usepackage{verbatim}
\usepackage{framed}
\usepackage{setspace}
\usepackage{cite}
\usepackage[tracking = true]{microtype}
\usepackage[T1]{fontenc}

\usepackage[textsize=footnotesize]{todonotes}
\presetkeys{todonotes}{fancyline}{}
\usepackage{cite}
\usepackage{algorithmic}
\usepackage{graphicx}
\usepackage{textcomp}
\usepackage{xcolor}
\usepackage{booktabs}
\usepackage{hyperref}
\usepackage{cleveref}
\usepackage{listings}
\usepackage{textcomp}
\usepackage{threeparttable}
\usepackage[english=american,babel=false]{csquotes}
\usepackage{adjustbox}
\usepackage{array}
\usepackage{booktabs}

\usepackage{ftnright}

\newif\ifabridged
\newif\ifnotabridged
\newif\ifanonymous
\newif\ifnotanonymous
\newif\ifreviewer
\newif\ifnotreviewer

\ifdefined\isabridged
\abridgedtrue
\fi

\ifabridged\notabridgedfalse
\else\notabridgedtrue
\fi

\ifdefined\isanonymous
\anonymoustrue
\fi

\ifanonymous\notanonymousfalse
\else\notanonymoustrue
\fi

\ifdefined\isreviewer
\reviewertrue
\fi

\ifreviewer\notreviewerfalse
\else\notreviewertrue
\fi

\newcommand{\dCOne}{\ding{192}\xspace}
\newcommand{\dCTwo}{\ding{193}\xspace}
\newcommand{\dCThree}{\ding{194}\xspace}
\newcommand{\dCFour}{\ding{195}\xspace}

\newcommand{\cmark}{\ding{51}}%
\newcommand{\xmark}{\ding{55}}%

\newcommand{\unicTWO}{\includegraphics[scale=0.35]{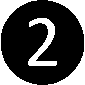}}
\newcommand{\unicTHREE}{\includegraphics[scale=0.35]{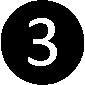}}
\newcommand{\unicFOUR}{\includegraphics[scale=0.35]{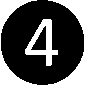}}
\newcommand{\unicFIVE}{\includegraphics[scale=0.35]{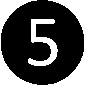}}
\newcommand{\unicSIX}{\includegraphics[scale=0.35]{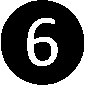}}
\newcommand{\unicSEVEN}{\includegraphics[scale=0.35]{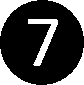}}
\newcommand{\unicEIGHT}{\includegraphics[scale=0.35]{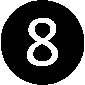}}
\newcommand{\unicNINE}{\includegraphics[scale=0.35]{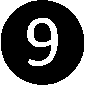}}
\newcommand{\unicTEN}{\includegraphics[scale=0.35]{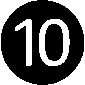}}
\newcommand{\unicELEVEN}{\includegraphics[scale=0.35]{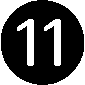}}
\newcommand{\unicTWELVE}{\includegraphics[scale=0.35]{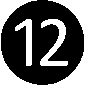}}
\newcommand{\unicTHIRTEEN}{\includegraphics[scale=0.35]{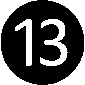}}

\DeclareFontEncoding{LS1}{}{}
\DeclareFontSubstitution{LS1}{stix2}{m}{n}
\DeclareRobustCommand{\boxonbox}{%
  \usefont{LS1}{stix2frak}{m}{n}\symbol{'270}%
}
\DeclareFontEncoding{LS2}{}{}
\DeclareFontSubstitution{LS2}{stix2}{m}{n}
\DeclareRobustCommand{\squareulquad}{%
  \usefont{LS2}{stix2tt}{m}{n}\symbol{'217}%
}

\newcommand{\SDRoB}{SDRaD\xspace}

\newcommand{\PKRUSAFE}{PKRU-S\textsc{afe}}
\newcommand{\TRUST}{TR\textsc{ust}}

\renewcommand{\paragraph}[1]{\noindent\textbf{#1.}}

\input{listing-rust.sty}

\newcolumntype{R}[2]{%
    >{\adjustbox{angle=#1,lap=\width-(#2)}\bgroup}%
    l%
    <{\egroup}%
}

\begin{document}

\title{Friend or Foe Inside? Exploring In-Process Isolation to Maintain
Memory Safety for Unsafe Rust}


\ifnotanonymous
\author{\IEEEauthorblockN{Merve G\"ulmez}
\IEEEauthorblockA{\textit{Ericsson Security Research} \\ Kista, Sweden \\
\textit{imec-DistriNet, KU Leuven} \\ Leuven, Belgium \\
merve.gulmez@kuleuven.be}
\and
\IEEEauthorblockN{Thomas Nyman}
\IEEEauthorblockA{\textit{Ericsson Product Security} \\ Jorvas, Finland \\
thomas.nyman\\@ericsson.com}
\and
\IEEEauthorblockN{Christoph Baumann}
\IEEEauthorblockA{\textit{Ericsson Security Research} \\
Kista, Sweden \\
christoph.baumann\\@ericsson.com}
\and
\IEEEauthorblockN{Jan Tobias M\"uhlberg}
\IEEEauthorblockA{%
\textit{imec-DistriNet, KU Leuven} \\ Leuven, Belgium \\
\textit{Universit\'e Libre de Bruxelles} \\ Brussels, Belgium \\
jan.tobias.muehlberg@ulb.be}
}
\fi

\maketitle


\begin{abstract}
Rust is a popular memory-safe systems programming language. In order to interact
with hardware or call into non-Rust libraries, Rust provides \emph{unsafe}
language features that shift responsibility for ensuring memory safety to the
developer. Failing to do so, may lead to memory safety violations in unsafe code
which can violate safety of the entire application. In this work we explore
in-process isolation with Memory Protection Keys as a mechanism to shield safe
program sections from safety violations that may happen in unsafe sections. Our
approach is easy to use and comprehensive as it prevents heap and stack-based
violations. We further compare process-based and in-process isolation mechanisms
and the necessary requirements for data serialization, communication, and
context switching.  Our results show that in-process isolation can be effective
and efficient, permits for a high degree of automation, and also enables a
notion of application rewinding where the safe program section may detect and
safely handle violations in unsafe code.

\end{abstract}



\section{Introduction}\label{sec:introduction}

Rust is an emerging system programing language with memory safety guarantees
and performance characteristics close to traditional system programming
languages such as C and C++~\cite{Pereira_2017}. Rust is designed to make
programs difficult to exploit by attackers by providing compile-time
type-checking and memory management based on ownership rules. These static
analyses in Rust are conservative: While the compiler will never accept an
unsafe program, it may reject safe programs. To work around this
incompleteness of compile-time analyses but also to accommodate the need to
interact with inherently unsafe low-level operating system interfaces or
hardware, Rust provides \emph{unsafe Rust}~\cite{Klabnik18}. In unsafe Rust
it is possible to, e.g.,  dereference a raw pointer or to modify a
mutable static variable. Rust then relies on the developer for the
correctness of unsafe parts of a program, and failure to ensure correctness
will lead to memory errors that may compromise safe parts of the program.

A common use of unsafe Rust is to call C library functions. While the Rust
ecosystem is growing, it is not feasible to reimplement common library
functionality in Rust right away and the Rust community embraced a Foreign
Function Interface (FFI,~\cite{Klabnik18}) to conveniently call non-Rust
code. Foreign functions are always assumed to be unsafe by the compiler and
it obliges the developer to integrate C libraries safely. Common causes
of unsafety regarding the FFI are library interfaces that are not
thread-safe, pointer arguments that are not completely sanity checked, and
the use of raw pointers. The use of FFI is pervasive: According to Li et
al.~\cite{Li2022} more than~72\% of packages on the official Rust package
registry (\url{crates.io}) depend on at least one unsafe FFI-bindings
package.
%
With many high-profile open-source projects such as the Mozilla
Gecko browser engine ($\approx$ 10\% Rust\footnote{Mozilla Gecko
repository: \url{https://github.com/mozilla/gecko-dev}}) adopting Rust,
ongoing efforts to add support for the language to the Linux
kernel\footnote{Rust for Linux project: \url{https://rust-for-linux.com/}},
and Microsoft announcing the uptake of Rust in the Windows
OS,\footnote{\enquote{You will actually have Windows booting with Rust in
the kernel in probably the next several weeks or months, which is really
cool.} -- David Weston, director of OS security for Windows, at BlueHat IL
2023.}
the language is gaining a strong foothold across industry sectors and
developing easy-to-use and efficient ways to safely integrate legacy 
libraries through FFI is highly relevant~\cite{Mergendahl2022}.

Earlier approaches exist to make invoking unsafe code safer, e.g., by executing
unsafe code in a separate process~\cite{Lamowski2017} or by utilizing Memory
Protection Keys (MPK) to provide in-process heap
isolation~\cite{Liu2020,Kirth22}, based either on developer knowledge or on
automated inference in a compilation framework~\cite{Bang23}.  Separating safe
and unsafe program sections into multiple processes has the advantage of
combining comprehensive heap and stack isolation with the potential of
recovering execution of the safe application part after a crash of the unsafe
process, albeit incurring substantial runtime overheads. In-process isolation,
in comparison, has mostly been used to isolate safe from unsafe heaps only and
with exceptions leading to program termination, yet with much smaller overheads.

\paragraph{This Paper and Contributions}
In this work we study Secure Rewind and Discard of Isolated
Domains~\cite{gulmez2022dsn} to protect Rust applications that make use of
unsafe language features. As a mechanism that enables compartmentalized
in-process isolation of safe and unsafe program sections, our approach
relies on MPK to provide stack and heap protection, and allows the safe
compartment to recover from violations caused in unsafe code, albeit
without relying on a custom compiler and with much smaller overheads than
earlier related work that achieves similar properties by means of
process-based isolation. Our study also engages with the question of how to
fairly compare different approaches to compartmentalization and isolation
as they require different approaches to memory management, data
serialization, and communication between compartments, which incur a
majority of the overheads. Specifically, we make the following
contributions:
\begin{itemize}
  \item{We present SDRaD-FFI to protect the integrity of a Rust application
from memory-safety violations in unsafe program parts or C libraries,
utilizing in-process isolation. SDRaD-FFI further increases the Rust
application's availability through a secure rewinding mechanism.}
  \item{We implement in-process isolation stack and heap protection for
Foreign Function Interface for Rust by leveraging the SDRaD C library. We
provide an easy-to-use API for SDRaD-FFI, which provides a high degree of
automation.}
  \item{We compare the costs of context switching and serialization for
process-based and in-process isolation using libpng and snappy libraries as
case studies. Specifically we evaluate different serialization approaches
in Rust.}
\end{itemize}
We aim to open-source our prototype and experimental evaluation data
upon publication of this paper.

We discuss the Rust FFI and approaches to isolating unsafe code in
Sect.~\ref{sec:background} and explain the objectives of our study in
Sect.~\ref{sec:problem_statement}. In Sect.~\ref{sec:implementation} and 
\ref{sec:evaluations} we detail SDRaD-FFI, and
experimentally compare our prototype with related approaches. We report on a
security evaluation, lessons learned, and limitations of our approach in
Sect.~\ref{sec:discussion}. Finally, in Sect.~\ref{sec:related_work} and
\ref{sec:conclusions}, we discuss our approach in the context of recent
related work and draw conclusions.

\section{Background}\label{sec:background}
\subsection{Rust-FFI}
\label{sec:rust-ffi}

Rust is a modern systems programming language that is designed to prevent
memory-safety defects through a strong type system combined with built-in
compile- and run-time checks.  Memory-safe Rust will restrict arbitrary casting,
prevent temporal memory-safety bugs through a set of ownership rules on data
objects, and perform bounds-checks on static and dynamic data allocations.
Together, these properties ensure \emph{safe} Rust code is \emph{sound} and will
not exhibit undefined behavior~\cite{rustonomicon}.

However, Rust programmers also have access to an \emph{unsafe} superset of Rust
which may exhibit undefined behavior.  With unsafe Rust it is the programmer
(not the compiler) who is responsible for ensuring the soundness of the code.
Unsafe Rust is generally required when interfacing directly with hardware,
operating systems, or \emph{other languages}.

\paragraph{Foreign function interface}
The Rust Foreign Function Interface (FFI) enables the sharing
of code and functions between Rust and other programming languages.  Since the
target language may not conform to the memory-safety properties enforced in
Rust, FFI calls are considered inherently unsafe.  As Rust and other memory-safe
system programming languages are being deployed gradually, it is frequently
necessary for Rust programs to interface with legacy libraries written in unsafe
languages, such as C and C++.  Recent work on the security of such
multi-language programs~\cite{Mergendahl2022} demonstrate that the interplay
between safe and unsafe languages can undermine existing mitigations against
memory attacks.

Different approaches for sandboxing legacy code from the safe Rust code
have been proposed~\cite{Almohri18, Lamowski2017, Liu2020, Kirth22}.  In what
follows, we describe the most relevant approaches in detail.

\subsection{Process-based Isolation}\label{sec:procisolation}

Process-based isolation is an essential concept in most operating systems.
It protects the system's integrity and resilience by providing the following features:
\begin{inparaenum}[1)]
    \item  \textit{integrity}: each process runs in its own virtual memory space that prevents
a malicious process from accessing the memory of another process, and
    \item  \textit{resilience:} each process has its own failure
boundary so one process' failure does not affect others.
\end{inparaenum}

\paragraph{Multi-process software architectures}
Compartmentalizing large applications into distinct, isolated processes is a well-known design pattern used both for security and reliability.
\ifnotabridged
For example, modern web browsers, such as Chrome\footnote{\url{https://developer.chrome.com/blog/inside-browser-part1/}}, compartmentalize different components, such as the browser chrome, renderer, plugins, and GPU tasks into distinct processes.
Typically, each different tab has at least one renderer process of its own.
This allows for resilience compared to running a common rended process for all tabs; if the renderer becomes unresponsive, all tabs are unresponsive.
In a multi-process renderer architecture when a tab becomes unresponsive other tabs can still function independently.
For security, recent versions of Chrome further compartmentalize the tab-specific renderer into site-specific renderers where each cross-site iframe is rendered in a separate, isolated process.
The sandboxing of renderers in such as manner that mimics the same-origin policy security model is called \emph{site isolation}~\cite{Reis19}.
\fi
Multi-process software architectures, such as \emph{site-isolating} browsers come with two significant drawbacks:
\begin{inparaenum}[1)]
    \item  they are complex to engineer and maintain, and
    \item  come with associated memory and process-to-process communication overhead that is exacerbated as the number of processes increases.
\end{inparaenum}

The engineering challenges with process-based isolation stem from moving from a monolithic program with a single thread of control to a concurrent programming model that necessitate multiple isolated processes to co-ordinate their execution and to communicate results to each other. For example, the incorporating of site isolation to the Chrome browser is the result of a multi-year engineering effort.

\paragraph{Automatic software compartmentalization}
The problem of \emph{automatically} splitting an existing monolithic program
into multiple, compartmentalized components requires program transformation of
local function calls into \emph{remote procedure calls} (RPCs) that occur across
multiple processes.  Sensible boundaries for such RPCs are highly
application-specific, e.g., for Chrome, they are defined by the interfaces
between the browser chrome and the isolated component, such as the renderer.  In
most prior work automatic software compartmentalization the boundary is based on
existing library APIs that, using source-to-source translation, can be turned
into RPC calls.  Source-to-source translation is generally invasive and requires
changes to the compiler which rarely make their way into upstream toolchains.

\paragraph{Sandcrust}
In this work, we primarily focus on the use case of automatically
compartmentalizing unsafe library interfaces in Rust programs.
Sandcrust~\cite{Lamowski2017} is an easy-to-use sandboxing solution for
compartmentalizing Rust applications to execute unsafe code in a separate,
isolated process.  Unlike solutions that require compiler-based source-to-source
translation Sandcrust builds on the metaprogramming capabilities provided by the
Rust macro system. This is the key selling point for Sandcrust's ease-of-use: to
compartmentalize functions belonging to an unsafe library interface, they are
simply annotated with a \texttt{sandbox!} macro provided by the Sandcrust crate
(\Cref{lst:sandcrust}).

\begin{lstlisting}[language=Rust, float, style=boxed, label = {lst:sandcrust}, caption = {Transform unsafe $F$ with Sandcrust's \texttt{sandbox!} macro to a remote procedure executed in an isolated process.}, captionpos=b]
sandbox!{
    fn F (...) {
        unsafe {
            ... // unsafe code
        }
    }
}
\end{lstlisting}

Under process-based isolation, for different compartments (i.e., processes) to
communicate and exchange data with each other, they must do so via
\emph{inter-process communication} (IPC) primitives.  IPC typically comes with
high overheads due to the context switching associated with scheduling different
process and data crossing security and failure boundaries.  In Sandcrust,
argument passing and communicating return values between the main process and
sandboxed code is handled over IPC channels.  To allow the passing of Rust (and
C) objects across the process boundary, one must be able to serialize, and
deserialize, any data type that is to be transferred.  As we will show in
\Cref{sec:serialization} the overhead associated with serialization and
deserialization is the principle source of run-time overhead for
compartmentalization in Sandcrust.  Consequently, minimizing the overhead of
object passing to and from the sandboxed code is an important consideration for
making automatic program compartmentalization in Rust more tractable.

\subsection{In-Process Isolation}\label{sec:inprocisolation}

In contrast to process-based isolation, \emph{in-process isolation} is based on
the notion of creating compartmentalized security domains within the memory
space of a single application process.  The main perceived benefit of in-process
isolation is that since transitions from one domain to another stays within the
same process, in-process isolation can significantly reduce the run-time cost of
context switching compared to traditional process isolation.  This is especially
beneficial when domain transitions are frequent.

\paragraph{Software fault isolation}
In-process isolation is enabled through \emph{software fault isolation}
(SFI)~\cite{Tan17}, a technique for using program transformations to establish
\emph{logical} protection domains.  The enforcement of such protection domains
can either leverage software-based checks inserted through compiler-based
program transformation~\cite{Wahbe93, Castro09, Mao11, PtrSplit}, binary
rewriting~\cite{Erlingsson06}, or hardware-assisted
enforcement~\cite{Rivera16,Koning17,Hedayati19,Vahldiek-Oberwagner19,Melara19,Sung20,Lefeuvre21,Schrammel20,Wang20,Voulimeneas22,Kirth22,Jin22,Chen22}.

\paragraph{Memory protection keys}
The inclusion of \emph{memory protection
keys} (MPK)~\cite{protectionkeys} to commodity processors has prompted advances in SFI solutions that
leverage hardware assistance.  The \emph{protection keys for userspace} (PKU)
mechanism in 64-bit x86 processors by both Intel~\cite{Intel64} and
AMD~\cite{AMD-AMD64} has received, by far, the most attention from academia.

\paragraph{Protection keys for userspace}
PKU associates each 4KB memory page with a 4-bit \emph{protection key} which is
stored by the operating system (OS) in the page table entry.  An
additional 32-bit, CPU-specific, user accessible, \emph{protection key rights
register} (PKRU) stores a 2-bit value for each of the 16 possible protection
keys, controlling whether associated memory pages are writable or accessible.
The validity of the memory accesses is enforced in hardware based on the PKRU
configuration.

Because the PKRU is accessible from user space, the access control policy
enforced by PKU can be updated without the need to call into the OS kernel.
This characteristic makes PKU significantly more efficient compared to
OS-enforced memory access control; OS-involvement is only needed when a memory
page's protection key is updated.  However, as the PKRU policy is entirely
controlled from user space, it can also be subverted by adversaries than can
control writes to the PKRU.  This risks violating the isolation guarantees of
PKU-enforced in-process isolation unless PKU is augmented with protection for
unauthorized writes to the PKRU.  Previous work has explored compiler-based code
rewriting~\cite{Koning17}, binary inspection~\cite{Vahldiek-Oberwagner19},
system call filtering~\cite{Voulimeneas22}, and hardware
extensions~\cite{Schrammel20} to harden PKU/PKRU security.

\paragraph{\PKRUSAFE{} and XRust}
Automatic compartmentalization of multi-language Rust applications using PKU has
been explored by Liu et al.~\cite{Liu2020} and Kirth et al.~\cite{Kirth22}.
These prior works focus on portioning the heap into distinct protection domains
for safe and unsafe Rust code, including calls occurring via FFIs.  While these
compartmentalization approaches share similar goals to process-based isolation,
they are not directly comparable as process-based isolation encompasses not only
the application heap, but all data, including the stack and static allocations.
As discussed in \Cref{sec:procisolation}, process-based isolation
can further provide a degree of resilience against memory corruption whereas
conventional SFI approaches terminate the application as soon as a violation of
a domain boundary is detected.

In this work, we are concerned with comparing process- and in-process isolation
for Rust-FFI under settings which provide similar isolation guarantees.
Consequently, \PKRUSAFE{} and XRust do not meet our requirements.

\paragraph{Secure rewind and discard}
Secure Rewind and Discard (SDRaD) of Isolated Domains~\cite{gulmez2022dsn} allows
compartmentalizing an application into distinct domains and restoring the
execution state of the application in case of memory corruption in the
compartmentalized
domain. \SDRoB
is a C library that realizes this scheme using in-process isolation based on
PKU.  Developers can enhance their applications with rewinding capability by
leveraging SDRaD APIs.  For example, a compartmentalized domain can be created
by assigning a custom user domain index with \texttt{sdrad\_init()} call. Memory
management for separate domain heaps is provided via \texttt{sdrad\_malloc()}
and \texttt{sdrad\_free()} calls, leveraging the TLSF
allocator~\cite{tlsf}. Developers can use \texttt{sdrad\_enter()} to enter and
run code in a previously initialized domain and \texttt{sdrad\_exit()} to exit
from the domain. In case a memory corrupting event is detected inside the
domain, the execution flows is rewound to the point where the domain was
initialized.  The application can then take an alternate action to avoid the
offending event.

Because restoring the process to a prior point of execution under an adversary
model where an attacker has access to the application memory requires strong
isolation guarantees that encompass both the application, stack, heap, and
static data, \SDRoB provides a stronger isolation model compared to
XRust~\cite{Liu2020} and \PKRUSAFE~\cite{Kirth22}.  Depending on the
configuration, the \SDRoB APIs can provide both confidentially and integrity
guarantees for the isolated domains.

The main of the drawbacks of \SDRoB is that manual effort is required for
integrating SDRaD API calls into an application and the current implementation only
supports C code.  Therefore, while \SDRoB provides a better match for the
isolation properties we require, it requires augmentation in order to be usable
with Rust code.

\section{Problem Statement}\label{sec:problem_statement}
The goal of this study is to apply in-process isolation as implemented by the
SDRaD library to the Rust FFI. Doing so, will allow developers to isolate
foreign functionality in an in-process sandbox with resilience to potential
memory safety violations. If such a fault occurs, the sandbox is discarded and
an error is signaled to developer-provided handler code that may either take
steps to recover from the fault (and avoid it to be triggered again, if
possible) or fail gracefully.

An important objective here is to use Rust's metaprogramming capabilities to
allow for automatic software compartmentalization in the same way as Sandcrust
does for process isolation. As discussed above, such ease-of-use is crucial for
the adoption of a hardening mechanism, as high complexity and cost has proven to
be one of the major obstacles to the improvement of software security.

The validation phase then compares the resulting prototype with Sandcrust as a
representative of automated process isolation for the Rust FFI. In particular,
we look at the performance overhead of the two solutions, micro-benchmarking the
context switch cost as well as evaluating real-world applications for different
data transfer volumes. Moreover we compare the security posture of the two
solutions. All study results are discussed in \Cref{sec:discussion} along with
lessons learned.

\section{Prototype Implementation}\label{sec:implementation}
In order to protect the Rust FFI using in-process isolation, we
provide \emph{SDRaD for Foreign Function Interfaces (SDRaD-FFI)}, a Rust crate
containing a Linux library for the 64-bit x86 architecture. It allows developers
to leverage metaprogramming in Rust to conveniently wrap functions that should
be executed in an isolated domain. Under the hood, it uses the \SDRoB C library
API with PKU as the underlying isolation primitive and it
supports compilation with different serialization crates.

\subsection{High Level Idea}\label{sec:highl_level}
As in \emph{Sandcrust}, our SDRaD-FFI crate provides a \texttt{sandbox!} macro to annotate functions that are to be isolated (cf.~\Cref{lst:sandcrust}).
Whenever a sandboxed function is called at run-time it executes in a nested
domain, i.e., isolated from the caller, using SDRaD.  In case of
rewinding after detecting a fault, the behavior depends on the return type of the
function. If Rust's \emph{Result}-type idiom~\cite{resulttype} is
used, the fault is encapsulated as an \emph{Err} variant.
Otherwise, SDRaD-FFI raises an unwinding \texttt{panic}~\cite{macropanic} that
can be caught using the standard library
\texttt{std::panic::catch\_unwind}
function~\cite{catchunwind} (cf.~\Cref{lst:rewinding}).
In either case the developer decides how to resume: they may try to recover,
e.g., by dropping the input that caused the fault, or exit the program
gracefully.

\begin{lstlisting}[language=Rust, float, style=boxed, label = {lst:rewinding}, caption = {Example code snippet demonstrating handling  unwinding panics raised by isolated function
$F$~\dCTwo. Function \texttt{std::panic::catch\_unwind} turns an
arbitrary function into a \emph{Result}-type function~\dCOne. The
result~\dCThree{} and error~\dCFour{} handling can be done as if
a \emph{Result} was directly returned.}, captionpos=b]
let result = std::panic::catch_unwind(||{ %*\dCOne*)
    F(...); %*\dCTwo*)
    }); 
match result { 
    Ok(v)  => ... , // Process result v %*\dCThree*)
    Err(e) => println!("Fault in nested domain!"), %*\dCFour*)
}
\end{lstlisting}


\begin{figure}[t]
    \centering
    \includegraphics[scale=0.34]{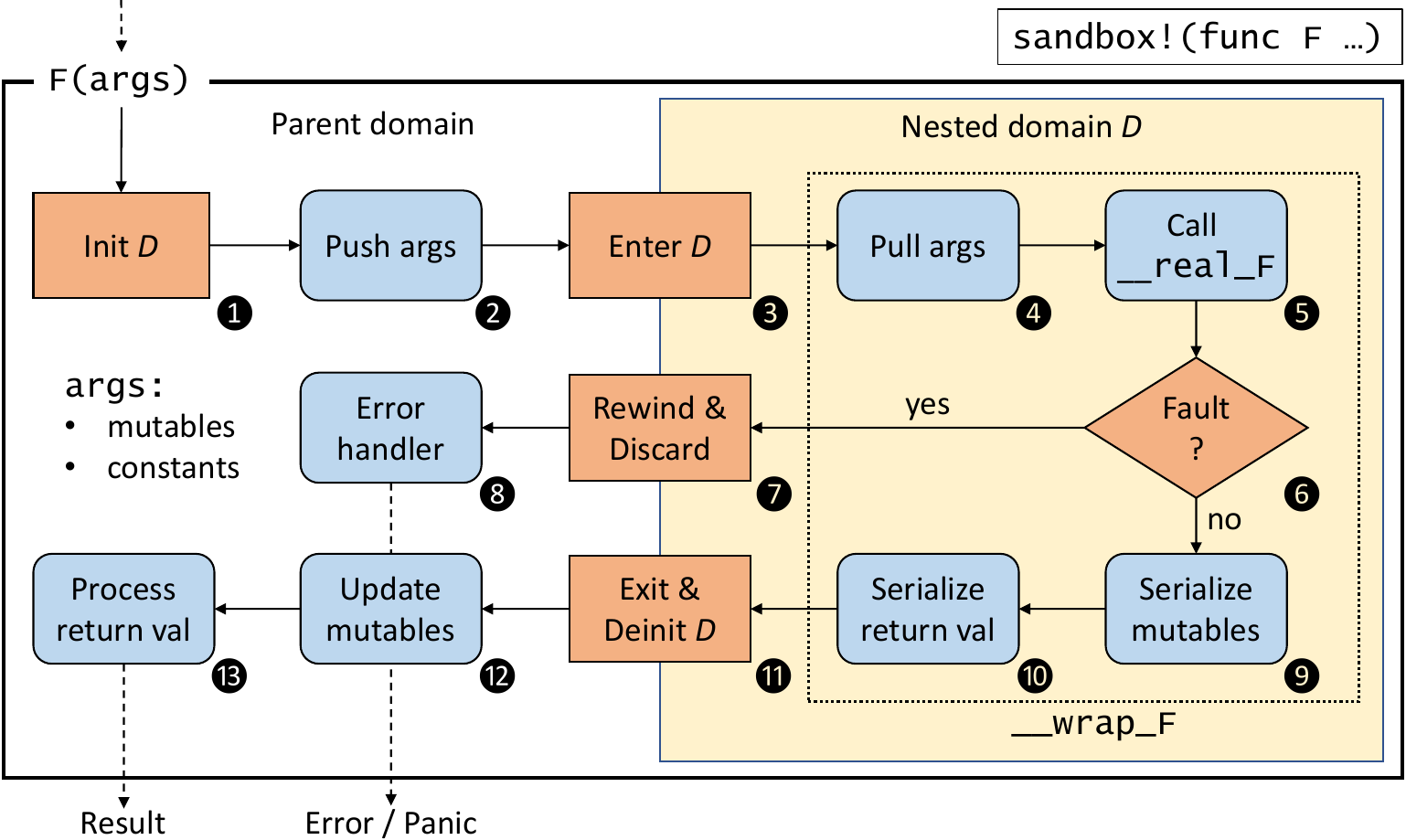}
    \caption{Transformation of sandbox macro. The angled orange boxed are SDRaD functionality, while rounded blue boxes are new glue logic for parameter passing and error handling.}
    \protect\label{fig:sdroblifecycle}
\end{figure}

\subsection{SDRaD-FFI design}
The \texttt{sandbox!} macro is expanded into calls to the \emph{SDRaD} library
and additional glue logic for passing parameters and results between the domains
as well as error handling. Syntactically, the definition of a sandboxed function
$F$ that takes a number of arguments and returns some value is maintained
as \texttt{\_\_real\_F()}, while an additional wrapper \texttt{\_\_wrap\_F()} is
defined to execute the required glue logic in the nested
domain. \Cref{fig:sdroblifecycle} shows the general control flow for executing a
call to \emph{F} in a nested domain $D$.

Firstly, domain $D$ is initialized as an accessible domain
via \texttt{sdrad\_init()} so that the parent domain can easily push the
arguments of $F$ into $D$'s memory. To this end, the parent domain serializes
the function arguments (cf.~\Cref{sec:serialization}) and writes them to a
dedicated memory area in the nested domain \unicTWO{} (cf.~\Cref{sec:ipc}).
Next, domain $D$ is entered using \texttt{sdrad\_enter()}~\unicTHREE{} and
wrapper \texttt{\_\_wrap\_F()} is executed in the nested domain area. It first
pulls from memory and rebuilds all function arguments using the deserialization
method~\unicFOUR{} (cf.~\Cref{sec:serialization}). Then, \texttt{\_\_real\_F()} holding the original function
body is executed~\unicFIVE.

If a memory access violates the nested domain boundaries~\unicSIX{} the fault is
caught by SDRaD, the domain is rewound and discarded automatically~\unicSEVEN{},
and control-flow is redirected to an error handler which either returns an error
value or raises a panic as described above~\unicEIGHT.\@
Similar to as in SDRaD for C code, also Rust applications can be compiled with stack
smashing protections\footnotemark{} that can trigger a domain rewind in
SDRaD-FFI.\@
\footnotetext{\texttt{RUSTFLAGS='-C stack-protector=strong}'~\cite{stackprotector}}

On a succesfull FFI call the results have to be persisted in the safe part of
the program.
To this end, any mutable local variable function arguments~\unicNINE{} and the
return value~\unicTEN{} are serialized and execution returns to the parent
domain via \texttt{sdrad\_exit\(\)}~\unicELEVEN.\@
During this step, the control state of domain $D$ is also deinitialized to
prevent it from being re-entered without a corresponding sandboxed call.
The parent domain then deserializes the mutable local variables, updates them in
the parent domain \unicTWELVE, and puts out the deserialized return value as a
result~\unicTHIRTEEN.\@

\subsection{In-Process Communication}\label{sec:ipc}
SDRaD-FFI requires the parent domain to manage a memory area in the nested
domain to transfer data between domains. SDRaD provides a fixed-size array
memory management API for nested domains, however, it requires require manual
size calculation and error handling, e.g., due to insufficient memory. To hide
these issues from the programmer, we leverage the Rust vector type, a
dynamically sized and resizable array, for data transfers between domains.
Rust vector allocation can be customized by defining specific Allocator
traits~\cite{traitallocator} for different use cases.
In our case, we need to create a new vector that should
grow and shrink using memory from a specific nested domain. To achieve it, we
implemented the Allocator trait for \emph{SdradAllocator} which is passed a
specific domain identifier to be used with the SDRaD memory management API.
Then a dedicated \emph{SdradAllocator} struct is initialized for each nested
domain and if arguments need to be passed from parent domain to a nested domain
$D$, a new vector is created for that purpose using $D$'s
dedicated \emph{SdradAllocator}.

During a call of a function $F$, all input data is copied to that vector by the
parent domain and read out by the nested domain when
executing \texttt{\_\_wrap\_F()}. To this end, the nested domain needs to know
vector metadata, i.e., the vector's length, capacity, and backend memory
information. The parent domain writes this metadata to a reserved region in the
nested domain stack. The nested domain can then create a second instance of the
same vector using \texttt{Vec::from\_raw\_parts\_in()} function by reading the
vector info from its own stack.

However, memory allocated for this vector in Rust is automatically freed when it
goes out of scope, i.e., when exiting \texttt{\_\_wrap\_F()}. As the original
instance of the vector references the same memory, that memory would be freed
once more when the parent function terminates eventually. To solve this problem,
we implemented an SDRaDNestedAllocator that does not free any memory area by
itself and we use it to construct the new vector in the nested memory area.

Some of the operations above are not natively supported
by the SDRaD API, which we extended (cf.~\Cref{sec:sdrad}).




\subsection{Serialization and Deserialization}\label{sec:serialization}
Before entering the newly spawned domain, the function arguments should be
passed and redefined in the nested domain.  To this end, it is required to track
the Rust data types of the arguments and copy all related memory areas into the
nested domain area.
Sandcrust~\cite{Lamowski2017} uses the Bincode serialization crate to pass
arguments to another process using IPC. Bincode transforms data into a common
binary representation that allows passing data between different platforms.
However, as Sandcrust and SDRAD-FFI target only a single platform, this is
redundant and Bincode serialization introduces unnecessary overhead. Sandcrust
provides a macro-level optimization for $\mathit{Vec}\langle \mathit{u8}\rangle$
to reduce this overhead.  It uses the macro matching mechanism to directly
serialize and deserialize data of type $\mathit{Vec}\langle \mathit{u8}\rangle$
by copying the vector memory area, sending it over the IPC, and later restoring
it in another process instead of using Bincode's native serialization scheme.
Unfortunately, this solution is not scalable, as it requires to provide a
dedicated optimized implementation for any other Rust type instance that is not
exactly $\mathit{Vec}\langle \mathit{u8}\rangle$, e.g.,
$\mathit{Vec}\langle \mathit{u16}\rangle$ or
$\mathit{Option}(\mathit{Vec}\langle \mathit{u8}\rangle)$.

Abomonation~\cite{abomonation} is
another Rust serialization crate based on memory layout representation. It does
not store any metadata or type system information in memory, and it can
deserialize in place without the need for another copy operation. That allows it
to be fast compared to other serialization crates\cite{rustsb}, but
not very suitable for typical serialization scenarios. Here, however, it is a
good fit, and we employ it as the basis of our (de)serialization
method. Another advantage of using Abomonation is that it
conveniently returns the remaining buffer after deserializing a variable,
removing the need for error-prone offset calculation and memory buffer
operations in the macro definitions when multiple arguments are passed.
In addition, the Rust specialization feature allows traits to have default
implementations that can be overridden for more specific types, allowing for
optimized trait implementations.~\cite{usingspecialization}.
We used this feature to extend the Abomonation trait, optimizing serialization
and deserialization of $\mathit{Vec}\langle \mathit{u8}\rangle$ in the same way as
Sandcrust does. As opposed to the Sandcrust implementation, using specialization
for traits is more scalable, as it also applies optimizations for sub-type
instances of a given type, e.g., for $\mathit{Vec}\langle \mathit{u8}\rangle$ within
$\mathit{option}(\mathit{Vec}\langle \mathit{u8}\rangle)$ in our case.

Our case studies also required passing arguments and results of
Rust's \emph{slice}
type, which
represents sections of vectors, strings, or arrays. Slices are not
a \emph{collection type}, i.e., their size is undefined at compile-time and
ownership of the underlying memory does not transfer with them, thus
deserialization into slices is usually not supported by Rust serialization
crates. We employed macro matching to solve this issue by first converting such
variables into vectors (for slices of vectors and arrays) or strings (for slices
of strings) before transferring them between domains using the above-mentioned
mechanism. After the transfer, each variable is cast back to its original slice
type.

\subsection{SDRaD Integration and Extension}\label{sec:sdrad}
In the SDRaD-FFI Rust crate, we create a binding to the SDRaD C library using
the $\#[link(...)]$ attribute. The SDRaD APIs are defined in an
$\texttt{extern}$ block in Rust, and we also define the SDRaD API macros
because the C library macro definitions cannot be
used directly in the Rust application.

In Rust, the default memory allocator used for dynamic memory management is the
system allocator provided by the operating system.  However, using alternative
allocators such as \emph{jemalloc} or \emph{tcmalloc} is also possible by
configuring the Rust compiler.  Supporting different allocators may have
advantages, such as reducing fragmentation for different applications.  However,
SDRaD uses the Two-Level Segregated Fit (TLSF) allocator~\cite{mattcontetlsf, tlsf} via interposing libc malloc family in the
library load order, which is necessary to allow the subheap memory management in
the application.  That restricts allowing different memory allocators in Rust.


As we discussed in \Cref{sec:ipc}, transferring data between domains requires
that the nested domain stack information should be exposed to the parent
domain. We extend the \SDRoB API with two new APIs:
\begin{inparaenum}[1)]
    \item exposing the stack base pointer to the caller, allowing the parent domain to write argument vector metadata in the nested domain stack area. 
    \item updating the stack base pointer of the nested domain that prevents overwriting information stored on the nested domain's stack by the parent domain. 
\end{inparaenum} 
These extensions were sufficient to support the in-process communication mechanism described above.

\subsection{Parallel and Nested Domains}\label{sec:nesting}
The SDRaD library allows to manage up 15 in-process domains using PKU. These
domains come in different flavors, e.g., accessible or inaccessible by the
parent domain as well as persistent or transient. Furthermore, different domains
can be used in parallel by different libraries and threads and even be nested to
achieve more fine-grained isolation.

We restrict the features of domains in favor of a simplified usage
model. By default, the nested domain to isolated foreign functionality from Rust
is accessible and persistent to simplify data passing and allow foreign
functionality with persistent state on the heap. The
developer may manage different domains by providing a custom \emph{Secure Domain
Identifier (SDI)} to easily separate persistent internal data,
e.g., two different libraries would use two
different SDIs in their \texttt{sandbox!} definitions.

In principle our implementation also supports multi-threading and nesting of
sandboxed calls, however, we consider such use cases out of scope of this study.

\section{Evaluation}\label{sec:evaluations} 

To evaluate our SDRaD-FFI prototype, we perform microbenchmarks to measure the latency of domain
transition via the SDRaD call gate and SDRaD heap allocator function call overheads. 
We evaluate the performance of SDRaD-FFI
applied to function calls from two unsafe C libraries: the compression library
\emph{snappy} and the image codec \emph{libpng}. 

We evaluated SDRaD-FFI performance using rustc
1.71.0-nightly, Abomonation 0.7.3, bincode 1.3.3, and 2.0.0-rc.3. 
We compared SDRaD-FFI with Sandcrust using bincode v1.0.0-alpha7. 
SDRaD-FFI leverages \texttt{allocator\_api} unstable features.
We run all experiments with the "release" build option on Dell PowerEdge R540 machines with
24-core MPK-enabled Intel(R) Xeon(R) Silver 4116 CPU (2.10GHz) having 128 GB RAM
and using Ubuntu 18.04, Linux Kernel 4.15.0.

\subsection{Microbenchmark}\label{sec:micro}
We perform microbenchmarks to measure the overhead of invoking a function in an
isolated domain and the overhead of the SDRaD version of \texttt{malloc()}
and \texttt{free()}. All benchmarks were conducted over $1\times 10^8$
iterations and we calculated the mean execution time and standard deviation
across all iterations.

\begin{figure}[t]
  \centering
  \includegraphics[scale=0.550]{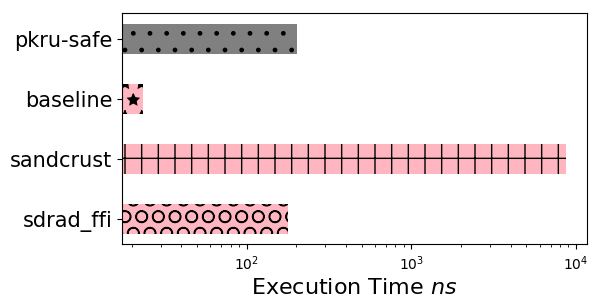}
  \caption{Microbenchmark of context switch latency for different sandboxing
    mechanisms. Numbers for PKRU-Safe have been extrapolated from literature.}
  \label{fig:micro}
\end{figure}

To gauge the overall context switch costs, we sandboxed
the \texttt{empty()} function and measured its execution time, i.e., the time
for entering and exiting the sandbox. We compared SDRaD-FFI to a baseline
value without compartmentalization, Sandcrust, and PKRU-Safe. Note that the
PKRU-Safe microbenchmark value is an estimate based on the description in the
paper~\cite{Kirth22}. Moreover, PKRU-Safe only compartmentalizes heap memory
space, and we report the number here as a reference for another PKU-based
isolation mechanism. \Cref{fig:micro} shows the execution times for all systems
under test. While the sandboxed \texttt{empty()} function with SDRaD-FFI takes
$177.2\mathit{ns}$ ($\sigma$=$61\mathit{ns}$), the baseline without sandboxing
takes $23\mathit{ns}$ ($\sigma$=$21\mathit{ns}$) and
Sandcrust takes $8671\mathit{ns}$ ($\sigma$=$695.5\mathit{ns}$). SDRaD-FFI is
$48.93x$ much faster than Sandcrust and $7.69x$ slower than the baseline
function calls. This result is in the same order of magnitude as context switch
overheads reported for PKRU-Safe~\cite{Kirth22}. 

We profiled the context switch and a main culprit for the overhead seems to be
CPU pipeline flushes due to writes of the PKRU register. In SDRaD, each API call
involves two such writes for entering and exiting the security monitor. Due to
initialization and deinitialization calls, entering and exiting the sandbox
accrues four PKRU writes each.

In the second case, we measured the execution time of calling \texttt{malloc()}
and \texttt{free()} functions by allocating and later freeing memory with sizes
uniformly distributed over a range from 0 to 4096 bytes.  We compare SDRaD with
the Rust default allocator. SDRaD takes $66\mathit{ns}$
($\sigma$=$40.5\mathit{ns}$) for \texttt{malloc()} and $48.1\mathit{ns}$
($\sigma$=$34.2\mathit{ns}$) for \texttt{free()} on average. In the Rust default
allocator, the average run time is $44.4\mathit{ns}$
($\sigma$=$31.1\mathit{ns}$), and $33.5\mathit{ns}$ ($\sigma$=$25.0\mathit{ns}$)
respectively. Thus, in the SDRaD allocator, \texttt{malloc()} is on average
$1.49x$ slower and \texttt{free()} is $1.44x$ slower compared to the Rust
default allocator.

\subsection{Snappy}
Snappy~\cite{snappy} is a fast compression
library that is presented as the FFI example in the Rust
Book~\cite{Klabnik18}, where
the raw snappy C APIs are wrapped by Rust interface functions:
\texttt{compress()} and \texttt{uncompress()} as seen in~\Cref{lst:snappy}.  We
sandboxed these functions with SDRaD-FFI, similar to
Sandcrust~\cite{Lamowski2017}.

Compressing and uncompressing randomly generated data of different sizes, we
measured the execution time of each operation for $5\times 10^5$ iterations.  We
evaluate SDRaD-FFI solutions with different serialization crates: bincode and
Abomonation.  Comparing \SDRoB-FFI to Sandcrust, we used the latter's "custom
vector" feature for optimized $\mathit{Vec}\langle u8\rangle$ serialization.
\begin{lstlisting}[language=Rust, float, style=boxed, label = {lst:snappy}, caption = {Snappy compress and uncompress function signature}]
  pub fn compress(src: &[u8]) -> Vec<u8>; 
  pub fn uncompress(src: &[u8]) -> Option<Vec<u8>>; 
\end{lstlisting}

\begin{figure*}[t]
  \begin{subfigure}[t]{0.48\textwidth}
    \centering
    \includegraphics[scale=0.300]{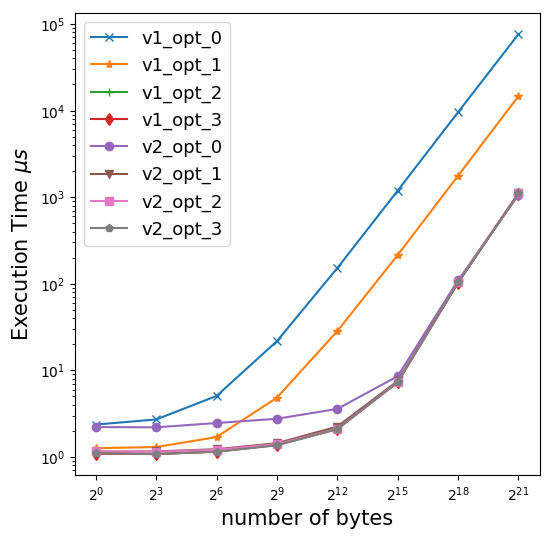}
    \includegraphics[scale=0.300]{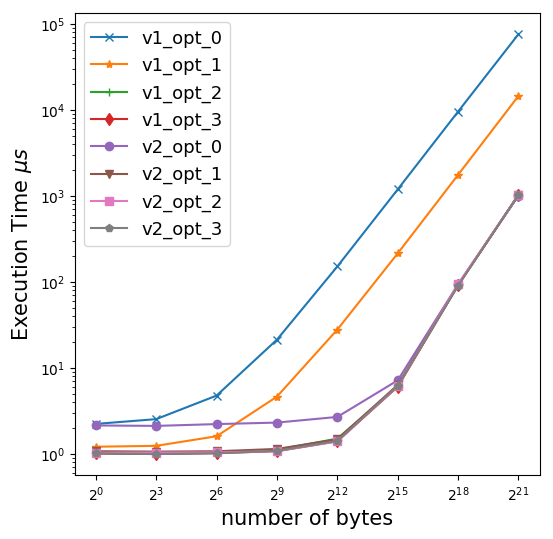}
    \caption{snappy \texttt{compress()} and \texttt{uncompress()} for SDRaD-FFI
    with Abomonation (v1: original, v2: specialized), compiled with different optimization levels}
    \label{fig:perf1_abomonation}
  \end{subfigure}
  \hfill
  \begin{subfigure}[t]{0.48\textwidth}
    \centering
    \includegraphics[scale=0.300]{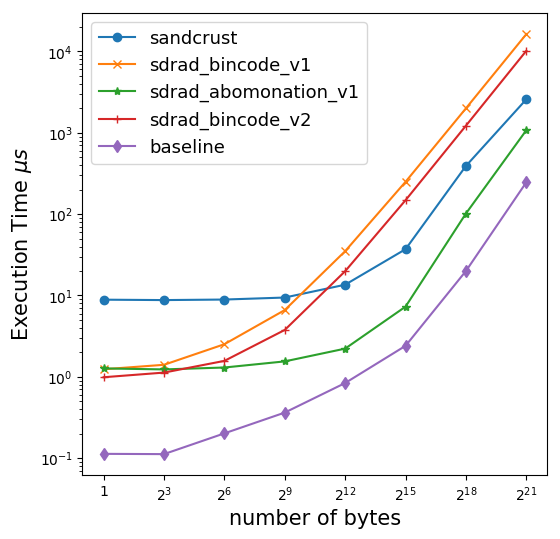}
    \includegraphics[scale=0.300]{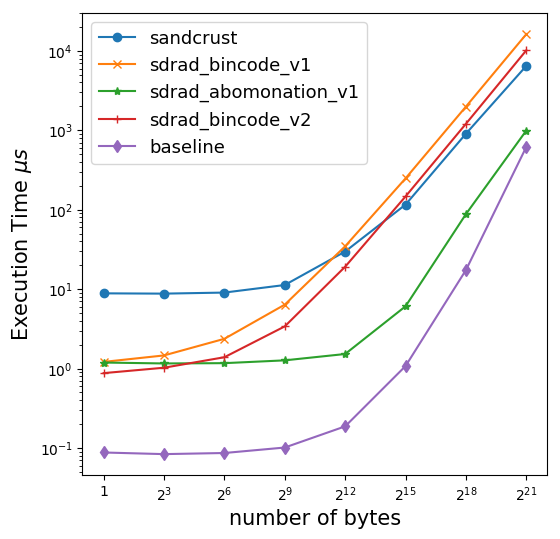}
    \caption{snappy \texttt{compress()} and \texttt{uncompress()} for SDRaD-FFI, Sandcrust, and baseline without any isolation}
    \label{fig:perf1_serialization}
  \end{subfigure}
  \caption{Execution Time of snappy with different isolation mechanisms for
  different numbers of bytes.}
  \label{fig:perf1}
\end{figure*}



\paragraph{Abomonation optimization} 
In the initial development phase of SDRaD-FFI, we noticed that Abomonation is
slow for $\mathit{Vec}\langle \mathit{T}\rangle$ type variables.  Abomonation
crate has a generic trait for $\mathit{Vec}\langle \mathit{T}\rangle$ variables
that are serialized and deserialized item by item and the corresponding iterator
code takes up around 60\% of the run time.  However, for
$\mathit{Vec}\langle \mathit{u8}\rangle$ such iteration is redundant: the vector
underlying memory area can simply copied in bulk.

To optimize Abomonation for $\mathit{Vec}\langle \mathit{u8}\rangle$, we used
the unstable specialization feature as explained \Cref{sec:serialization}.  Our
experiments show that this specialization improved performance significantly as
seen in \Cref{fig:perf1_abomonation}.

Later, we observed that building with optimization level 3 (release mode) yields
similar performance for the unmodified Abomonation crate with generic type
$\mathit{Vec}\langle \mathit{T}\rangle$. The Rust compiler optimizes the iteration
for $\mathit{Vec}\langle \mathit{u8}\rangle$ and inlines Abomonation completely
so that each serialization and deserialization step is just a
\texttt{memcpy()} operation.

Overall, as described in \Cref{sec:serialization}, the memory layout
presentation used by Abomonation to serialize and deserialize variables is
suitable for optimization, unlike Bincode, which uses binary format
representation and cannot be optimized automatically by the compiler in the same
way.

\paragraph{Snappy benchmark result} 
We perform the following benchmarking using the original version of Abomonation.
\Cref{fig:perf1} show the snappy \texttt{compress()} and \texttt{uncompress()}
function execution times for different byte sizes: 1 byte to $2^{21}$ bytes.
In \Cref{fig:perf1_serialization}, Sandcrust performs worst when compressing data
smaller than $2^{12}$ bytes which can be explained by our microbenchmark results
from~\Cref{sec:micro}.

For data larger than $2^{12}$ bytes, Sandcrust performs better than SDRaD-FFI
with bincode version bincode 1.3.3 and 2.0.0-rc.3. As discussed, the bincode serialization crate is
not optimized for serializing and deserializing byte arrays and Sandcrust uses
macro-level optimizations for $\mathit{Vec}\langle u8\rangle$ instead of using
the bincode serialization.  As a result, serialization crates overheads for
large bytes are dominant over overheads from isolation methods. SDRaD-FFI with
Abomonation crates is faster than the bincode-based versions and Sandcrust.

In general, all compartmentalization methods have an overhead compared to the
baseline of at least one order of magnitude if only little data is transferred
between the compartments. The overhead is reduced for larger data, but still
significant. We profiled the SDRaD-FFI benchmark using the \textit{perf}
tool and found that about 60\% of the run time is spent in \texttt{memcpy()}.

For uncompressing, Sandcrust has slightly better performance than SDRaD-FFI with
bincode 2.0.0-rc.3.  Even though Sandcrust is optimized to handle variables of
type $\mathit{Vec}\langle u8\rangle$, that solution does not apply for other
types that contain $\mathit{Vec}\langle u8\rangle$.  Sandcrust cannot perform
the same optimization for the \texttt{uncompress()} return value of type
$\mathit{Option}\langle\mathit{Vec}\langle u8\rangle\rangle$.

SDRaD-FFI with Abomonation does not suffer from this problem by integrating optimizations in the serialization mechanism directly. 
Its performance comes relatively close to the baseline for large inputs but degrades for small inputs due to the high context switch costs.

\subsection{Libpng}
The C library \emph{libpng} is used to handle portable network graphic images.
To apply SDRaD-FFI for the C library, we sandboxed libpng in an example of a
Rust application similar to the Sandcrust code published
in~\cite{lamowski2017automatic}. The example code reads the images and performs
PNG image decoding operations using libpng.  We compared
SDRaD-FFI-Abomonation with Sandcrust and measured the execution time of each
decoding operation for $1\times 10^6$ iterations and different image sizes.

\Cref{tab:png_measurements} shows mean and relative-standard deviation for decoding measurement of different images. 
SDRaD-FFI introduced a worst-case overhead of $11.72\%$ performance degradation compared to a baseline measurement for $5.5K$ image decoding and it performed significantly better than Sandcrust in all cases.

\begin{table*}[h]
    \begin{tabular}{|c|c|c|c|c|c|c|c|}
      \hline 
      &\multicolumn{3}{c|}{Execution Time [$\mu{}s$]} & \multicolumn{2}{c|}{Execution Time Overhead}\\ 
      Data size & \textbf{Baseline} & \textbf{SDRaD-FFI} & \textbf{Sandcrust} & \textbf{SDRaD-FFI} \ &  \textbf{Sandcrust} \ \\\ 
      (bytes)   &                   &                    &                   &  vs.~\textbf{Baseline}          &   vs.~\textbf{Baseline}        \\ 
      \hline
      5.5K      &    264.14  ($\sigma$$\pm$3.30\%)         &       295.10 ($\sigma$$\pm$5.33\%)      &    461.86 ($\sigma$$\pm$3.30\%)     &    +11.72\%          &     +74.85\%             \\
      64K       &   4430.93  ($\sigma$$\pm$4.00\%)        &       4749.53 ($\sigma$$\pm$3.19\%)      &   7694.12 ($\sigma$$\pm$2.91\%)     &    +7.19\%           &     +73.64\%             \\
      380K      &   9085.24  ($\sigma$$\pm$3.16\%)         &      9295.79 ($\sigma$$\pm$2.14\%)      &  11375.59 ($\sigma$$\pm$2.50\%)     &    +2.32\%           &     +25.21\%             \\
      895K      &  13415.76  ($\sigma$$\pm$3.39\%)         &     14008.36 ($\sigma$$\pm$2.00\%)      &  20513.90 ($\sigma$$\pm$2.81\%)     &    +4.42\%           &     +52.91\%             \\
      \hline
    \end{tabular}
    \centering
    \caption{\emph{libpng} Decoding Measurements}
    \label{tab:png_measurements}
  \end{table*}

\subsection{Security Evaluation}
We reproduced CVE-2018-1000810~\cite{CVE-2018-1000810} to verify the SDRaD-FFI error handling mechanisms. 
The \texttt{str::repeat} function in the Rust standard library has an integer overflow that causes a buffer overflow. 
Integer overflows only occur in the release build; Rust checks for these in
debug builds. 
We sandboxed the \texttt{str::repeat} function with a specific return value in case a fault occurs and the sandbox is discarded. 
We found that the buffer overflow causes a domain violation which triggers the rewinding mechanism. Control is transferred to the error handler which returns the specific return value.

\section{Discussion}\label{sec:discussion}\label{sec:extensions}\label{sec:applicability}
Below we summarize the insights gained in our study, provide a
comparative security
evaluation of both process-based and in-process isolation, and
outline 
remaining challenges.

\subsection{Lessons Learned}
As we have seen, the in-process isolation approach clearly outperforms process
isolation in terms of context switch overhead. This is not surprising, as the
development of PKU technology is explicitly designed to reduce context switch
cost (be it for switches to the kernel or to other user processes).

Nevertheless, even for modestly sized arguments, the context switch cost starts
to get dominated by the cost of data transfer between domains. Here, the data
serialization method used can significantly impact performance and thus it is
crucial to optimize it for the use case at hand.

One important insight here is that in our scenario, where data is exchanged
between isolated code running on the same platform, serialization can
essentially be simplified to copying the required data between different memory
domains and making sure that data is interpreted using the correct types.

Such an optimization is straight-forward for vector-type variables, however,
care must be taken where to implement these optimizations. We found that it is
more elegant and scalable to use traits and specializations within the
serialization crate instead of implementing optimizations using macro type
matching, as is done in Sandcrust.

In this study, we adapted in-process isolation as provided by SDRaD, which
involves many options for domains and a rather complicated
programming model. We strived to control this complexity 
and provide a simple interface for isolation of
unsafe code, similar to Sandcrust.

The Rust macro system proved to be a powerful tool in this respect as its
meta-programming capabilities allow to automate argument and result
passing. Moreover, the \emph{Result} type and \emph{panic} concepts could
readily be integrated with the sandbox design. If possible, unsafe functions
should use the \emph{Result} return type idiom, so that they can easily be
sandboxed without the developer having to write a custom panic handler.

Prior work on automated software compartmentalization conventionally relies on
compiler-based source-to-source translation which, as discussed in
\Cref{sec:procisolation}, faces a high barrier for adoption due to requiring
invasive changes to compilers and toolchains.  Compartmentalization APIs such as
\SDRoB place the burden of integrating compartmentalization capability on the
developers.  For languages such as C, such changes can become invasive as the
amount of data to be passed across the domain boundaries increases.  Languages
with strong metaprogramming capabilities, such as the Rust macro system can
greatly reduce the complexity for repetitive integration effort, such as
argument passing.  There can also be benefits in converging on interoperable
interfaces for language- and library-level compartmentalizing; by adopting a
similar interface as Sandcrust, SDRaD-FFI can be used as a drop-in replacement
for Sandcrust that provides improved run-time performance.


\subsection{Security Evaluation}

The security of Sandcrust is based on the mutual isolation of regular user
processes. This is a fundamental and well-studied concept of computer systems~\cite{Daley68, Saltzer75, Karger02}
and a multitude of language, compiler, and OS-based hardening techniques exists
to raise the bar for adversaries to breach the confines of a user
process.~\cite{Aiken06, Kosaka18, Schwarzl21}

In comparison, hardware-assisted in-process isolation has only been widely
supported since rather recently and is not without caveats. In particular,
the PKU mechanism as implemented by Intel and AMD has known security issues that
necessitate additional compile-time effort and runtime hardening to ensure that
the in-process isolation is watertight. 

The required measures have been
discussed before in detail~\cite{gulmez2022dsn,Connor20,Schrammel22,Voulimeneas22}, in a
nutshell:
\begin{inparaenum}
\item instrumented programs need to be scanned for illicit writes to the PKRU register which controls access to domain memory~\cite{Koning17,Vahldiek-Oberwagner19,Connor20},
\item a control flow integrity mechanism needs to be in place to ensure that legitimate PKRU write instructions of the security monitor cannot be abused as gadgets by an adversary~\cite{Hedayati19,Connor20},
\item\label{point:modifycode}run-time modification of code must either be prohibited or subject to binary analysis to verify that no new PKRU write instructions are introduced~\cite{Vahldiek-Oberwagner19},
\item additional system call filtering and hardening of signal handlers is
needed to prevent abuse by an adversary who tampers with the PKU mechanism~\cite{Connor20,Voulimeneas22,Schrammel22}.
\end{inparaenum}
While these are not unsurmountable issues, especially requirement~\ref{point:modifycode}) makes it tenuous
to support functionality that requires just-in-time compilation.

However, improvements of the PKU hardware design have been suggested which would
make most of the hardening measures listed above redundant~\cite{Schrammel20}. It would be desirable for
processor designers such as Intel and AMD to adopt similar solutions to make
in-process isolation more secure and usable.

\subsection{Future Work}

As a direct consequence of our work, the performance of Sandcrust could be
improved by employing a more streamlined serialization crate such as
Abomonation.
As we
have shown, process isolation is inherently slower than in-process isolation due
to the context switch cost, which is especially pronounced for low transfer
bandwidth between the safe and unsafe parts of the program. Hence we shift focus
to in-process isolation.

As discussed before, SDRaD-FFI can natively support parallel and nested
isolation of foreign functionality. Future work should investigate which use
cases could benefit from these features as well as their performance impact. 

Another avenue for improvement is the security of SDRaD-FFI itself. As it is
based on the SDRaD C library, our SDRaD-FFI crate is mostly unsafe code itself
which increases the TCB. Formal verification could be
employed to obtain correctness and security guarantees on the in-process
isolation implementation.

One may ask whether SDRaD-FFI can only be used for the Foreign Function
Interface or also to isolate unsafe functionality in general. This interface is
a natural place to split a program into two, as there is likely no
interdependency between functionality written in different languages and it is
convenient to split the program stack at a function call.
Yet, for general unsafe code that tends to be integrated
into the control flow within a Rust function, working on shared objects in
memory is essential. Here it becomes more challenging to split a Rust program into two
domains without breaking the Rust runtime and its control-flow integrity and memory safety
guarantees.

Moreover, unsafe code in Rust is often used to optimize operations for low
latency in ways that would not be legal in plain Rust~\cite{Klabnik18}. Given the significant
performance overhead of in-process isolation, it might actually be cheaper to
use a safe Rust implementation than to isolate an unsafe code snippet. However,
certain performance optimization might not be easy to achieve without the use of
unsafe Rust~\cite{rustonomicon}. Further research is needed to identify classes
of "native" unsafe Rust code for which in-process isolation is practical. In
some cases, the Rust standard library may need to be augmented to support
execution confined by an isolated domain~\cite{Liu2020}.


\section{Related Work}\label{sec:related_work}
\newcommand*\rotate{\multicolumn{1}{R{30}{1em}}}%
\newcommand*\leftalign{\multicolumn{2}{l}}%
\newcommand*\rightalign{&}%
\newcommand{\tablenote}[1]{\footnotesize{#1}}

\newcommand{\process}{{\boxonbox}\xspace}
\newcommand{\inprocess}{{\squareulquad}\xspace}

\newcommand{\yes}{\cmark\xspace}
\newcommand{\no}{\xmark\xspace}
\newcommand{\na}[1][\xspace]{\multirow{2}{*}{--#1}}

\begin{table*}
    \caption{Comparison of application compartmentalization schemes discussed in \Cref{sec:related_work}. The first rows shows how many distinct, isolated protection domains are supported. The \emph{isolation type} is either process (\process) or in-process (\inprocess) isolation. The next seven rows indicate what kind of code modules may be isolated w.r.t.~stack and heap memory. \emph{Crash resistance} means that the application can continue execution after protection domain violations are detected and contained. We characterize the \emph{development effort} as follows: the scheme requires invasive code changes, e.g., modifying the arguments of functions (\emph{medium}), superficial code changes suffice (\emph{low}), or the scheme can be incorporated \emph{automatically}. The last rows show the geometric mean of the \emph{performance degradation} for \TRUST~\cite{Bang23}, Sandcrust\cite{Lamowski2017} and SDRaD-FFI in the Snappy benchmark (\Cref{sec:evaluations}) compress and uncompress operations for input sizes 256B, 1KB, 4KB, 16KB, 64K, 256K, and 1GB as used by Bang et al.~\cite{Bang23}.}
    \begin{minipage}[t]{\textwidth}
    \begin{center}
    \resizebox{\textwidth-5em}{!}{
    \begin{threeparttable}
    \begin{tabular}{ l l c c c c c c c c c }
                                                            &
                                                            &
        \rotate{\textbf{ERIM}~\cite{Vahldiek-Oberwagner19}} &
        \rotate{\textbf{SDRaD}~\cite{gulmez2022dsn}}        &
        \rotate{\textbf{XRust}~\cite{Liu2020}}              &
        \rotate{\textbf{Fidelius Charm}~\cite{Almohri18}}   &
        \rotate{\textbf{Galeed}~\cite{Rivera21}}            &
        \rotate{\textbf{\PKRUSAFE}~\cite{Kirth22}}          &
        \rotate{\textbf{\TRUST}~\cite{Bang23}}              &
        \rotate{\textbf{Sandcrust}~\cite{Lamowski2017}}     &
        \rotate{\textbf{SDRaD-FFI}}                         \\
        \toprule
        \leftalign{No. of domains}      & 16\tnote{1}  & 15\tnote{1}  & 2           & 2           & 2           & 2           & 2           & 2           & 15\tnote{1} \\
        \leftalign{Isolation type}      & \inprocess{} & \inprocess{} & \inprocess  & \inprocess  & \inprocess  & \inprocess  & \inprocess  & \process    & \inprocess  \\
        \cmidrule(lr){1-11}
        Between safe    & Isolated stack    & \na{\tnote{2}} & \na{\tnote{2}}  & \no{}  & \yes{}      & \no{}       & \no{}       & \yes{}      & \yes{}      & \yes{}      \\
        and unsafe code & Isolated heap     &             &               & \yes{}      & \no{}       & \no{}       & \yes{}      & \yes{}      & \yes{}      & \yes{}      \\
        \cmidrule(lr){1-11}
        Between unsafe  & Isolated stack    & \yes{}       & \yes{}       & \no{}       & \no{}       & \no{}       & \no{}       & \no{}       & \yes{}\tnote{3}     & \yes{}      \\
        and unsafe code & Isolated heap     & \yes{}       & \yes{}       & \no{}       & \no{}       & \no{}       & \no{}       & \no{}       & \yes{}\tnote{3}      & \yes{}      \\
        \cmidrule(lr){1-11}
        \leftalign{Mixed-language support}  & \no{}    & \no{}        & \no{}       & \yes{}      & \yes{}      & \yes{}      & \yes{}      & \yes{}      & \yes{}      \\
        \cmidrule(lr){1-11}
        Between safe and & Isolated stack   & \na{}        & \na{}       & \na{}       & \yes{}       & \no{}       & \no{}       & \yes{}      & \yes{}      & \yes{}      \\
        mixed-language code & Isolated heap &              &             &             & \no{}        & \yes{}      & \yes{}      & \yes{}      & \yes{}      & \yes{}      \\
        \cmidrule(lr){1-11}
        \leftalign{Crash resistance}    & \no{}        & \yes{}      & \no{}       & \no{}        & \no{}       & \no{}       & \no{}       & \yes{}      & \yes{}      \\
        \leftalign{Development effort}  & med.         & med.        & low         & med.         & med.        & auto.       & auto.       & low         & low         \\
        \cmidrule(lr){1-11}
        Performance degradation & Compress & \multicolumn{6}{|c|}{\multirow{2}{*}{\textit{Numbers not available}}} & 20.5\%\tnote{4} &  1265.70\% &  155.7\%  \\
        in Snappy benchmark   & Uncompress & \multicolumn{6}{|c|}{                                               } & 50.0\%\tnote{4} & 6340.2\% &  370.8\%  \\
    \bottomrule
    \end{tabular}
    \begin{tablenotes}
    \item[1] \tablenote{May be less than $16 / 15$ in cases where Linux's MPK support reserves a few domains for specific purposes. \textsuperscript{2} Only supports C and / or C++ code.}
    \item[3] \tablenote{Due to manipulating the Rust abstract syntax tree (AST) via Rust's macros Sandcrust's sandbox can also be applied to unsafe functions without FFI.}
    \item[4] \tablenote{Result based on numbers reported in Table $8$ in pre-print of Bang et al.~\cite{Bang23} available at time of writing. We limit the comparison to the 256B -- 1GB range due to the custom allocator used by SDRaD limiting the size of continuous allocations to slightly under 4GB in size~\cite{gulmez2022dsn}.}
    \end{tablenotes}
    \end{threeparttable}
    }
    \end{center}
    \end{minipage}
    \label{tab:related_work}
\end{table*}

Several studies have explored the compartmentalizing of applications based on
different underlying primitives: hardware-assisted in-process
isolation~\cite{Rivera16,Koning17,Hedayati19,Vahldiek-Oberwagner19,Melara19,Sung20,Lefeuvre21,Schrammel20,Wang20,Voulimeneas22,Kirth22,Jin22,Chen22,
  gulmez2022dsn}, process-based isolation~\cite{Lamowski2017} and software-based
isolation~\cite{Tan17, Liu2020}.  The majority of such approaches address the
problem of compartmentalizing applications written in C or
C++~\cite{Koning17,Hedayati19,Vahldiek-Oberwagner19,Melara19,Sung20,Lefeuvre21,Schrammel20,Wang20,Voulimeneas22,Jin22,Chen22,
  gulmez2022dsn}.  In this work, we focus on multi-language applications,
specifically Rust code calling C or C++ code~\cite{Almohri18, Bang23, Kirth22,
  Lamowski2017, Liu2020}.  As noted in \Cref{sec:background}, the majority of
prior work on using in-process isolation operate under weaker isolation
guarantees than those of comparable solutions based on
process-isolation~\cite{Lamowski2017}; XRust~\cite{Liu2020},
Galeed~\cite{Rivera21} and \PKRUSAFE~\cite{Kirth22} only protect the Rust heap
from unsafe code while Fidelius Charm~\cite{Almohri18} only protects the Rust
stack.  

Independently and concurrently with our work, Bang et al.~\cite{Bang23} propose
\TRUST, a scheme with similar goals as Sandcrust~\cite{Lamowski2017}.  \TRUST{}
protects both the Rust heap and stack from unsafe and mixed-language code.
However, unlike SDRaD-FFI, which we designed to act as a drop-in replacement for
Sandcrust, \TRUST{} relies on compiler-driven compartmentalization of Rust code
based on the language-level boundary between safe and unsafe Rust.  The
principal benefit of compiler-based approaches, such as \PKRUSAFE{} and \TRUST{}
is relieving the developer completely of the burden of defining protection
domains.  In practice, automated approaches are restricted to boundaries that
can be derived from language-level constructs.  In the case of Rust, approaches
such as \TRUST{} are not suitable for isolating unsafe code from other unsafe
code.  Combined with conservative compartmentalization policies that classify
any objects that the untrusted code might use as unsafe, this can result in
problematic edge cases where nearly all allocations are delegated to the unsafe
domain~(cf.~Table 10 in \cite{Bang23}), negating the benefit of
compartmentalization in the first place.  Approaches, such as Sandcrust and
SDRaD-FFI, which \emph{minimize} developer effort but leave the developer in
control can deal efficiently with such cases.  Fully automating
compartmentalization using compiler-based rewriting can also, at least in the
short term, be counter-productive for the adoption of these solutions as
invasive changes to mature toolchains are required, which are unlikely to make
their way upstream.
Point-solutions that can be deployed as discrete libraries or crates offer a
more realistic path to industry adoption.  Finally, as discussed in
\Cref{sec:background}, one of the principal motivations for process-based
isolation are improved resilience by means of a failure boundary between
processes.  Apart from SDRaD and SDRaD-FFI, no prior work in in-process
isolation provides such a failure boundary between compartmentalized domains.

\Cref{tab:related_work} summarizes our comparison between the relevant
compartmentalization schemes for commodity 64-bit x86 processors excluding
approaches that require kernel modification~\cite{Almohri18, Hedayati19,
  Vahldiek-Oberwagner19} for deployment.  SDRaD-FFI is the only in-process
isolation solution that provides similar capabilities as process-based isolation
while significantly improving the run-time overhead compared to the similar
Sandcrust in the snappy use case. While SDRaD-FFI remains less efficient than
in-process isolation with weaker security guarantees such as \TRUST, we point
out that the measured performance overhead is highly benchmark-specific.  We
posit that schemes such as SDRaD-FFI remain a valid alternative for
compiler-based rewriting for use cases that require more flexibility than static
analysis can achieve to take developer intent into account.



\section{Conclusions}\label{sec:conclusions}
By allowing to include unsafe code, the Foreign Function Interface (FFI)
represents a chink in the armor of Rust's memory safety
guarantees~\cite{Almohri18, Lamowski2017, Liu2020, Kirth22, Mergendahl2022}. A
line of recent research tries to protect safe Rust code from the fallout of
potential memory safety violations in cross-language functionality, e.g., by
running such code as a separate process that may be compromised without
affecting the safety of the Rust code. In this paper we have studied the use of
in-process stack and heap memory isolation to secure the Rust FFI.

Building on the SDRaD C library~\cite{gulmez2022dsn} for secure domain
rewind and discard, we developed a prototype, SDRaD-FFI, which allows to
conveniently sandbox foreign functionality in Rust. The implementation
hides domain operations as well as passing arguments and results between
the domains using Rust's macro metaprogramming capabilities. Faults within
the isolated code are exposed to the programmer using standard error
handling idioms such as \emph{panics} and \emph{Result} types.

We compare our prototype to Sandcrust's implementation of process isolation for
FFI~\cite{Lamowski2017} and find that in-process isolation outperforms the
latter due to its lower context-switching costs. This is further improved by
optimized serialization methods for data transfer between isolated domains.  We
also compared the security of in-process isolation to the traditional
process-based approach and find that in-process isolation requires quite
extensive hardening to protect the PKU isolation mechanism. However, these
shortcomings can be alleviated by updates to the hardware design that have been
proposed in related work.

Overall, we conclude that in-process isolation is a usable and efficient
security solution for Rust FFI, especially when a lot of data is exchanged
between the safe and unsafe portions of the program. Our approach allows to
run unsafe foreign code alongside safe Rust code while isolating the latter
from possible memory safety violations in the former.
We aim to make our prototype and experimental evaluation data available
under an open-source license upon publication of this paper.

\ifnotanonymous
\section*{Acknowledgments} 
This work has received funding under EU H2020 MSCA-ITN action 5GhOSTS, grant agreement
no. 814035, by the Research Fund KU Leuven, by the
Flemish Research Programme Cybersecurity, and by the CyberExcellence
programme of the Walloon Region, Belgium.
\fi

\bibliographystyle{IEEEtran}
\bibliography{main}

\begin{thebibliography}{10}
\providecommand{\url}[1]{#1}
\csname url@samestyle\endcsname
\providecommand{\newblock}{\relax}
\providecommand{\bibinfo}[2]{#2}
\providecommand{\BIBentrySTDinterwordspacing}{\spaceskip=0pt\relax}
\providecommand{\BIBentryALTinterwordstretchfactor}{4}
\providecommand{\BIBentryALTinterwordspacing}{\spaceskip=\fontdimen2\font plus
\BIBentryALTinterwordstretchfactor\fontdimen3\font minus
  \fontdimen4\font\relax}
\providecommand{\BIBforeignlanguage}[2]{{%
\expandafter\ifx\csname l@#1\endcsname\relax
\typeout{** WARNING: IEEEtran.bst: No hyphenation pattern has been}%
\typeout{** loaded for the language `#1'. Using the pattern for}%
\typeout{** the default language instead.}%
\else
\language=\csname l@#1\endcsname
\fi
#2}}
\providecommand{\BIBdecl}{\relax}
\BIBdecl

\bibitem{Pereira_2017}
\BIBentryALTinterwordspacing
R.~Pereira, M.~Couto, F.~Ribeiro, R.~Rua, J.~Cunha, J.~a.~P. Fernandes, and
  J.~a. Saraiva, ``Energy efficiency across programming languages: How do
  energy, time, and memory relate?'' in \emph{Proceedings of the 10th ACM
  SIGPLAN International Conference on Software Language Engineering}, ser. SLE
  2017.\hskip 1em plus 0.5em minus 0.4em\relax New York, NY, USA: Association
  for Computing Machinery, 2017, p. 256–267. [Online]. Available:
  \url{https://doi.org/10.1145/3136014.3136031}
\BIBentrySTDinterwordspacing

\bibitem{Klabnik18}
S.~Klabnik and C.~Nichols, \emph{The Rust Programming Language}.\hskip 1em plus
  0.5em minus 0.4em\relax USA: No Starch Press, 2018.

\bibitem{Li2022}
Z.~Li, J.~Wang, M.~Sun, and J.~C.~S. Lui, ``Detecting cross-language memory
  management issues in rust,'' in \emph{Computer Security -- ESORICS 2022},
  V.~Atluri, R.~Di~Pietro, C.~D. Jensen, and W.~Meng, Eds.\hskip 1em plus 0.5em
  minus 0.4em\relax Cham: Springer Nature Switzerland, 2022, pp. 680--700.

\bibitem{Mergendahl2022}
S.~Mergendahl, N.~Burow, and H.~Okhravi, ``Cross-language attacks,'' 01 2022.

\bibitem{Lamowski2017}
\BIBentryALTinterwordspacing
B.~Lamowski, C.~Weinhold, A.~Lackorzynski, and H.~H\"{a}rtig, ``Sandcrust:
  {A}utomatic sandboxing of unsafe components in {R}ust,'' in \emph{Proceedings
  of the 9th Workshop on Programming Languages and Operating Systems}, ser.
  PLOS'17.\hskip 1em plus 0.5em minus 0.4em\relax New York, NY, USA:
  Association for Computing Machinery, 2017, p. 51–57. [Online]. Available:
  \url{https://doi.org/10.1145/3144555.3144562}
\BIBentrySTDinterwordspacing

\bibitem{Liu2020}
\BIBentryALTinterwordspacing
P.~Liu, G.~Zhao, and J.~Huang, ``Securing unsafe {Rust} programs with
  {XRust},'' in \emph{Proceedings of the ACM/IEEE 42nd International Conference
  on Software Engineering}, ser. ICSE '20.\hskip 1em plus 0.5em minus
  0.4em\relax New York, NY, USA: Association for Computing Machinery, 2020, p.
  234–245. [Online]. Available: \url{https://doi.org/10.1145/3377811.3380325}
\BIBentrySTDinterwordspacing

\bibitem{Kirth22}
\BIBentryALTinterwordspacing
P.~Kirth, M.~Dickerson, S.~Crane, P.~Larsen, A.~Dabrowski, D.~Gens, Y.~Na,
  S.~Volckaert, and M.~Franz, ``{PKRU-Safe}: Automatically locking down the
  heap between safe and unsafe languages,'' in \emph{Proceedings of the
  Seventeenth European Conference on Computer Systems}, ser. EuroSys '22.\hskip
  1em plus 0.5em minus 0.4em\relax New York, NY, USA: Association for Computing
  Machinery, 2022, p. 132–148. [Online]. Available:
  \url{https://doi.org/10.1145/3492321.3519582}
\BIBentrySTDinterwordspacing

\bibitem{Bang23}
\BIBentryALTinterwordspacing
I.~Bang, M.~Mayondo, H.~Moon, and Y.~Paek, ``{TR\textsc{ust}}: A compilation
  framework for in-process isolation to protect safe rust against untrusted
  code,'' in \emph{32nd {USENIX} Security Symposium ({USENIX} Security
  23)}.\hskip 1em plus 0.5em minus 0.4em\relax Baltimore, MD: {USENIX}
  Association, Aug. 2023. [Online]. Available:
  \url{https://www.usenix.org/conference/usenixsecurity23/presentation/bang}
\BIBentrySTDinterwordspacing

\bibitem{gulmez2022dsn}
M.~G\"ulmez, T.~Nyman, C.~Baumann, and J.~T. M\"uhlberg, ``Rewind \& {D}iscard:
  Improving software resilience using isolated domains,'' {To} appear in
  DSN'23, 2023, a technical report is available at
  \url{https://arxiv.org/pdf/2205.03205.pdf}.

\bibitem{rustonomicon}
{Rust Team}, ``{The Rustonomicon —- The Dark Arts of Advanced and Unsafe Rust
  Programming},'' Retrieved May 22, 2023 from
  \url{https://doc.rust-lang.org/nomicon/}.

\bibitem{Almohri18}
\BIBentryALTinterwordspacing
H.~M.~J. Almohri and D.~Evans, ``Fidelius charm: Isolating unsafe rust code,''
  in \emph{Proceedings of the Eighth ACM Conference on Data and Application
  Security and Privacy}, ser. CODASPY '18.\hskip 1em plus 0.5em minus
  0.4em\relax New York, NY, USA: Association for Computing Machinery, 2018, p.
  248–255. [Online]. Available: \url{https://doi.org/10.1145/3176258.3176330}
\BIBentrySTDinterwordspacing

\bibitem{Tan17}
\BIBentryALTinterwordspacing
G.~Tan, \emph{Principles and Implementation Techniques of Software-Based Fault
  Isolation}.\hskip 1em plus 0.5em minus 0.4em\relax Hanover, MA, USA: Now
  Publishers Inc., 2017. [Online]. Available:
  \url{https://www.cse.psu.edu/~gxt29/papers/sfi-final.pdf}
\BIBentrySTDinterwordspacing

\bibitem{Wahbe93}
\BIBentryALTinterwordspacing
R.~Wahbe, S.~Lucco, T.~E. Anderson, and S.~L. Graham, ``Efficient
  software-based fault isolation,'' in \emph{Proceedings of the Fourteenth ACM
  Symposium on Operating Systems Principles}, ser. SOSP '93.\hskip 1em plus
  0.5em minus 0.4em\relax New York, NY, USA: ACM, 1993, pp. 203--216. [Online].
  Available: \url{http://doi.acm.org/10.1145/168619.168635}
\BIBentrySTDinterwordspacing

\bibitem{Castro09}
\BIBentryALTinterwordspacing
M.~Castro, M.~Costa, J.-P. Martin, M.~Peinado, P.~Akritidis, A.~Donnelly,
  P.~Barham, and R.~Black, ``Fast byte-granularity software fault isolation,''
  in \emph{Proceedings of the ACM SIGOPS 22nd Symposium on Operating Systems
  Principles}, ser. SOSP '09.\hskip 1em plus 0.5em minus 0.4em\relax New York,
  NY, USA: ACM, 2009, pp. 45--58. [Online]. Available:
  \url{http://doi.acm.org/10.1145/1629575.1629581}
\BIBentrySTDinterwordspacing

\bibitem{Mao11}
\BIBentryALTinterwordspacing
Y.~Mao, H.~Chen, D.~Zhou, X.~Wang, N.~Zeldovich, and M.~F. Kaashoek, ``Software
  fault isolation with api integrity and multi-principal modules,'' in
  \emph{Proceedings of the Twenty-Third ACM Symposium on Operating Systems
  Principles}, ser. SOSP '11.\hskip 1em plus 0.5em minus 0.4em\relax New York,
  NY, USA: ACM, 2011, pp. 115--128. [Online]. Available:
  \url{http://doi.acm.org/10.1145/2043556.2043568}
\BIBentrySTDinterwordspacing

\bibitem{PtrSplit}
\BIBentryALTinterwordspacing
S.~Liu, G.~Tan, and T.~Jaeger, ``{PtrSplit: Supporting} general pointers in
  automatic program partitioning,'' in \emph{Proceedings of the 24th {ACM}
  Conference on Computer and Communications Security (ACM CCS)}, 2017.
  [Online]. Available: \url{https://dl.acm.org/doi/pdf/10.1145/3133956.3134066}
\BIBentrySTDinterwordspacing

\bibitem{Erlingsson06}
\BIBentryALTinterwordspacing
U.~Erlingsson, M.~Abadi, M.~Vrable, M.~Budiu, and G.~C. Necula, ``{XFI}:
  Software guards for system address spaces,'' in \emph{Proceedings of the 7th
  Symposium on Operating Systems Design and Implementation}, ser. OSDI
  '06.\hskip 1em plus 0.5em minus 0.4em\relax Berkeley, CA, USA: USENIX
  Association, 2006, pp. 75--88. [Online]. Available:
  \url{http://dl.acm.org/citation.cfm?id=1298455.1298463}
\BIBentrySTDinterwordspacing

\bibitem{Rivera16}
\BIBentryALTinterwordspacing
E.~E. Rivera, ``Preserving memory safety in safe rust during interactions with
  unsafe languages,'' Master's thesis, Department of Electrical Engineering and
  Computer Science, 2016. [Online]. Available:
  \url{https://dspace.mit.edu/bitstream/handle/1721.1/139052/Rivera-eerivera-meng-eecs-2021-thesis.pdf?sequence=1&isAllowed=y}
\BIBentrySTDinterwordspacing

\bibitem{Koning17}
\BIBentryALTinterwordspacing
K.~Koning, X.~Chen, H.~Bos, C.~Giuffrida, and E.~Athanasopoulos, ``No need to
  hide: Protecting safe regions on commodity hardware,'' in \emph{Proceedings
  of the Twelfth European Conference on Computer Systems}, ser. EuroSys
  '17.\hskip 1em plus 0.5em minus 0.4em\relax New York, NY, USA: Association
  for Computing Machinery, 2017, p. 437–452. [Online]. Available:
  \url{https://doi.org/10.1145/3064176.3064217}
\BIBentrySTDinterwordspacing

\bibitem{Hedayati19}
\BIBentryALTinterwordspacing
M.~Hedayati, S.~Gravani, E.~Johnson, J.~Criswell, M.~L. Scott, K.~Shen, and
  M.~Marty, ``Hodor: Intra-process isolation for high-throughput data plane
  libraries,'' in \emph{2019 {USENIX} Annual Technical Conference ({USENIX}
  {ATC} 19)}.\hskip 1em plus 0.5em minus 0.4em\relax Renton, WA: {USENIX}
  Association, Jul. 2019, pp. 489--504. [Online]. Available:
  \url{https://www.usenix.org/conference/atc19/presentation/hedayati-hodor}
\BIBentrySTDinterwordspacing

\bibitem{Vahldiek-Oberwagner19}
\BIBentryALTinterwordspacing
A.~Vahldiek-Oberwagner, E.~Elnikety, N.~O. Duarte, M.~Sammler, P.~Druschel, and
  D.~Garg, ``{ERIM}: Secure, efficient in-process isolation with protection
  keys ({MPK}),'' in \emph{28th {USENIX} Security Symposium ({USENIX} Security
  19)}.\hskip 1em plus 0.5em minus 0.4em\relax Santa Clara, CA: {USENIX}
  Association, Aug. 2019, pp. 1221--1238. [Online]. Available:
  \url{https://www.usenix.org/conference/usenixsecurity19/presentation/vahldiek-oberwagner}
\BIBentrySTDinterwordspacing

\bibitem{Melara19}
\BIBentryALTinterwordspacing
M.~S. Melara, M.~J. Freedman, and M.~Bowman, ``Enclavedom: Privilege separation
  for large-tcb applications in trusted execution environments,'' {\tt
  arXiv:1907.13245 [cs.CR]}, 2019. [Online]. Available:
  \url{http://arxiv.org/abs/1907.13245}
\BIBentrySTDinterwordspacing

\bibitem{Sung20}
\BIBentryALTinterwordspacing
M.~Sung, P.~Olivier, S.~Lankes, and B.~Ravindran, ``Intra-unikernel isolation
  with intel memory protection keys,'' in \emph{Proceedings of the 16th ACM
  SIGPLAN/SIGOPS International Conference on Virtual Execution Environments},
  ser. VEE '20.\hskip 1em plus 0.5em minus 0.4em\relax New York, NY, USA:
  Association for Computing Machinery, 2020, p. 143–156. [Online]. Available:
  \url{https://doi.org/10.1145/3381052.3381326}
\BIBentrySTDinterwordspacing

\bibitem{Lefeuvre21}
\BIBentryALTinterwordspacing
H.~Lefeuvre, V.-A. B\u{a}doiu, c.~Teodorescu, P.~Olivier, T.~Mosnoi,
  R.~Deaconescu, F.~Huici, and C.~Raiciu, ``Flexos: Making os isolation
  flexible,'' in \emph{Proceedings of the Workshop on Hot Topics in Operating
  Systems}, ser. HotOS '21.\hskip 1em plus 0.5em minus 0.4em\relax New York,
  NY, USA: Association for Computing Machinery, 2021, p. 79–87. [Online].
  Available: \url{https://doi.org/10.1145/3458336.3465292}
\BIBentrySTDinterwordspacing

\bibitem{Schrammel20}
\BIBentryALTinterwordspacing
D.~Schrammel, S.~Weiser, S.~Steinegger, M.~Schwarzl, M.~Schwarz, S.~Mangard,
  and D.~Gruss, ``Donky: Domain keys {\textendash} efficient in-process
  isolation for risc-v and x86,'' in \emph{29th {USENIX} Security Symposium
  ({USENIX} Security 20)}.\hskip 1em plus 0.5em minus 0.4em\relax {USENIX}
  Association, Aug. 2020, pp. 1677--1694. [Online]. Available:
  \url{https://www.usenix.org/conference/usenixsecurity20/presentation/schrammel}
\BIBentrySTDinterwordspacing

\bibitem{Wang20}
\BIBentryALTinterwordspacing
X.~Wang, S.~Yeoh, P.~Olivier, and B.~Ravindran, ``Secure and efficient
  in-process monitor (and library) protection with intel mpk,'' in
  \emph{Proceedings of the 13th European Workshop on Systems Security}, ser.
  EuroSec '20.\hskip 1em plus 0.5em minus 0.4em\relax New York, NY, USA:
  Association for Computing Machinery, 2020, p. 7–12. [Online]. Available:
  \url{https://doi.org/10.1145/3380786.3391398}
\BIBentrySTDinterwordspacing

\bibitem{Voulimeneas22}
\BIBentryALTinterwordspacing
A.~Voulimeneas, J.~Vinck, R.~Mechelinck, and S.~Volckaert, ``You shall not
  (by)pass! practical, secure, and fast pku-based sandboxing,'' in
  \emph{Proceedings of the Seventeenth European Conference on Computer
  Systems}, ser. EuroSys '22.\hskip 1em plus 0.5em minus 0.4em\relax New York,
  NY, USA: Association for Computing Machinery, 2022, p. 266–282. [Online].
  Available: \url{https://doi.org/10.1145/3492321.3519560}
\BIBentrySTDinterwordspacing

\bibitem{Jin22}
\BIBentryALTinterwordspacing
X.~Jin, X.~Xiao, S.~Jia, W.~Gao, H.~Zhang, D.~Gu, S.~Ma, Z.~Qian, and J.~Li,
  ``Annotating, tracking, and protecting cryptographic secrets with
  cryptompk,'' in \emph{2022 2022 IEEE Symposium on Security and Privacy (SP)
  (SP)}.\hskip 1em plus 0.5em minus 0.4em\relax Los Alamitos, CA, USA: IEEE
  Computer Society, may 2022, pp. 473--488. [Online]. Available:
  \url{https://doi.ieeecomputersociety.org/10.1109/SP46214.2022.00028}
\BIBentrySTDinterwordspacing

\bibitem{Chen22}
\BIBentryALTinterwordspacing
Y.~Chen, J.~Li, G.~Xu, Y.~Zhou, Z.~Wang, C.~Wang, and K.~Ren, ``{SGXLock}:
  Towards efficiently establishing mutual distrust between host application and
  enclave for {SGX},'' in \emph{31st USENIX Security Symposium (USENIX Security
  22)}.\hskip 1em plus 0.5em minus 0.4em\relax Boston, MA: USENIX Association,
  Aug. 2022. [Online]. Available:
  \url{https://www.usenix.org/conference/usenixsecurity22/presentation/chen-yuan}
\BIBentrySTDinterwordspacing

\bibitem{protectionkeys}
T.~kernel~development community, ``{Memory Protection Keys},'' Retrieved June
  08, 2023 from
  \url{https://www.kernel.org/doc/html/next/core-api/protection-keys.html}.

\bibitem{Intel64}
\emph{Intel 64 and IA-32 Architectures Software Developer's Manual Volume 3A:
  System Programming Guide}, Intel Corporation, 2007, order Number:
  325462-076US
  \url{https://www.intel.com/content/www/us/en/developer/articles/technical/intel-sdm.html}.

\bibitem{AMD-AMD64}
\emph{{AMD64 Architecture Programmer's Manual Volume 2: System Programming.
  Revision 3.38}}, {AMD}, 2021, publication No. 24593
  \url{https://www.amd.com/system/files/TechDocs/24593.pdf}.

\bibitem{tlsf}
\BIBentryALTinterwordspacing
M.~Masmano, I.~Ripoll, A.~Crespo, and J.~Real, ``Tlsf: a new dynamic memory
  allocator for real-time systems,'' in \emph{Proceedings. 16th Euromicro
  Conference on Real-Time Systems, 2004. ECRTS 2004.}, 2004, pp. 79--88.
  [Online]. Available: \url{https://ieeexplore.ieee.org/document/1311009}
\BIBentrySTDinterwordspacing

\bibitem{resulttype}
{The Rust Standard Library}, ``{Module std::result},'' Retrieved June 08, 2023
  from \url{https://doc.rust-lang.org/std/result/}.

\bibitem{macropanic}
------, ``{Macro std::panic},'' Retrieved June 08, 2023 from
  \url{https://doc.rust-lang.org/std/macro.panic.html}.

\bibitem{catchunwind}
------, ``{Function std::panic::catch\_unwind},'' Retrieved June 08, 2023 from
  \url{https://doc.rust-lang.org/std/panic/fn.catch\_unwind.html}.

\bibitem{stackprotector}
{Rust Team}, ``{GitHub rust-lang/rust pull request \#84197: add codegen option
  for using LLVM stack smash protection},'' Retrieved June 08, 2023 from
  \url{https://github.com/rust-lang/rust/pull/84197}.

\bibitem{traitallocator}
{The Rust Standard Library}, ``{Trait std::alloc::Allocator},'' Retrieved June
  08, 2023 from \url{https://doc.rust-lang.org/std/alloc/trait.Allocator.html}.

\bibitem{abomonation}
T.~Dataflow, ``{Abomonation},'' Retrieved June 08, 2023 from
  \url{https://github.com/TimelyDataflow/abomonation}.

\bibitem{rustsb}
D.~Koloski, ``{Rust Serialization Benchmark},'' Retrieved June 08, 2023 from
  \url{https://github.com/djkoloski/rust_serialization_benchmark#rust-serialization-benchmark}.

\bibitem{usingspecialization}
{The Rust Standard Library Developers Guide}, ``{Using spcialization},''
  Retrieved June 08, 2023 from
  \url{https://std-dev-guide.rust-lang.org/code-considerations/using-unstable-lang/specialization.html}.

\bibitem{mattcontetlsf}
M.~Conte, ``Github - mattconte/tlsf: Two-level segregated fit memory allocator
  implementation,'' April 2016, retrieved April 22, 2022 from
  \url{https://github.com/mattconte/tlsf}.

\bibitem{snappy}
Google, ``{Snappy},'' Retrieved June 08, 2023 from
  \url{https://github.com/google/snappy}.

\bibitem{lamowski2017automatic}
\BIBentryALTinterwordspacing
B.~Lamowski, ``Automatic sandboxing of unsafe software components in high level
  languages,'' 2017. [Online]. Available:
  \url{https://lamowski.net/docs/Automatic_Sandboxing_of_Unsafe_Software_Components_in_High_Level_Languages.pdf}
\BIBentrySTDinterwordspacing

\bibitem{CVE-2018-1000810}
CVE-2018-1000810, August 2018, retrieved March 09, 2023 from
  \url{https://nvd.nist.gov/vuln/detail/CVE-2018-1000810}.

\bibitem{Daley68}
\BIBentryALTinterwordspacing
R.~C. Daley and J.~B. Dennis, ``Virtual memory, processes, and sharing in
  multics,'' \emph{Commun. ACM}, vol.~11, no.~5, p. 306–312, may 1968.
  [Online]. Available: \url{https://doi.org/10.1145/363095.363139}
\BIBentrySTDinterwordspacing

\bibitem{Saltzer75}
J.~Saltzer and M.~Schroeder, ``The protection of information in computer
  systems,'' \emph{Proceedings of the IEEE}, vol.~63, no.~9, pp. 1278--1308,
  1975.

\bibitem{Karger02}
P.~A. Karger and R.~R. Schell, ``Thirty years later: Lessons from the multics
  security evaluation,'' in \emph{Proceedings of the 18th Annual Computer
  Security Applications Conference}, ser. ACSAC '02.\hskip 1em plus 0.5em minus
  0.4em\relax USA: IEEE Computer Society, 2002, p. 119.

\bibitem{Aiken06}
\BIBentryALTinterwordspacing
M.~Aiken, M.~F\"{a}hndrich, C.~Hawblitzel, G.~Hunt, and J.~Larus,
  ``Deconstructing process isolation,'' in \emph{Proceedings of the 2006
  Workshop on Memory System Performance and Correctness}, ser. MSPC '06.\hskip
  1em plus 0.5em minus 0.4em\relax New York, NY, USA: Association for Computing
  Machinery, 2006, p. 1–10. [Online]. Available:
  \url{https://doi.org/10.1145/1178597.1178599}
\BIBentrySTDinterwordspacing

\bibitem{Kosaka18}
M.~Kosaka, ``{Inside look at modern web browser (part 1)},'' Retrieved June 08,
  2023 from \url{https://developer.chrome.com/blog/inside-browser-part1/}.

\bibitem{Schwarzl21}
\BIBentryALTinterwordspacing
M.~Schwarzl, P.~Borrello, A.~Kogler, K.~Varda, T.~Schuster, D.~Gruss, and
  M.~Schwarz, ``Dynamic process isolation,'' {\tt arXiv: 2110.04751 [cs.CR]},
  2021. [Online]. Available: \url{https://arxiv.org/abs/2110.04751}
\BIBentrySTDinterwordspacing

\bibitem{Connor20}
\BIBentryALTinterwordspacing
R.~J. Connor, T.~McDaniel, J.~M. Smith, and M.~Schuchard, ``{PKU} pitfalls:
  Attacks on {PKU-based} memory isolation systems,'' in \emph{29th USENIX
  Security Symposium (USENIX Security 20)}.\hskip 1em plus 0.5em minus
  0.4em\relax USENIX Association, Aug. 2020, pp. 1409--1426. [Online].
  Available:
  \url{https://www.usenix.org/conference/usenixsecurity20/presentation/connor}
\BIBentrySTDinterwordspacing

\bibitem{Schrammel22}
\BIBentryALTinterwordspacing
D.~Schrammel, S.~Weiser, R.~Sadek, and S.~Mangard, ``Jenny: Securing syscalls
  for {PKU-based} memory isolation systems,'' in \emph{31st USENIX Security
  Symposium (USENIX Security 22)}.\hskip 1em plus 0.5em minus 0.4em\relax
  Boston, MA: USENIX Association, Aug. 2022. [Online]. Available:
  \url{https://www.usenix.org/conference/usenixsecurity22/presentation/schrammel}
\BIBentrySTDinterwordspacing

\bibitem{Rivera21}
\BIBentryALTinterwordspacing
E.~Rivera, S.~Mergendahl, H.~Shrobe, H.~Okhravi, and N.~Burow, ``Keeping safe
  rust safe with galeed,'' in \emph{Annual Computer Security Applications
  Conference}, ser. ACSAC '21.\hskip 1em plus 0.5em minus 0.4em\relax New York,
  NY, USA: Association for Computing Machinery, 2021, p. 824–836. [Online].
  Available: \url{https://doi.org/10.1145/3485832.3485903}
\BIBentrySTDinterwordspacing

\end{thebibliography}

\section{Appendix}

As supplemental material, we provide all measurements for results reported in
the this report.

\Cref{tbl:abomonation} shows execution times for running \emph{snappy} with
different versions of Abomonation under different optmization levels for
different input sizes. These are the numbers underlying
\Cref{fig:perf1_abomonation}.

Similarly, the numbers underlying the comparison of SDRaD-FFI with different
serialization crates and Sandcrust w.r.t.~performance of compression and
uncompression with \emph{snappy} as shown in \Cref{fig:perf1_serialization} are
presented in \Cref{tbl:snappy}.

Finally, we show the evaluation results for the comparison with \TRUST{} and
Sandcrust in \Cref{tab:related_work}. As the evaluation of \TRUST{} uses
different input sizes for \emph{snappy} we ran a separate experiment using the
sizes from Figure 10 in~\cite{Bang23}, except for 4GiB and 16GiB that where too
large to be allocated contiguously by the SDRaD library. We evaluated the
run-time of the baseline and SDRaD-FFI with Abomonation over 5000 iterations for
each input size. As Sandcrust runs incredibly slow for large input sizes, we
only measured each input size 50 times at the cost of a larger standard
deviation.

The resulting execution times from our experiments are depicted in
\Cref{fig:trust_compression} based on the raw numbers shown in
\Cref{tbl:trust}. We calculated the geometric mean of the measured run times and
those given in~\cite{Bang23} and computed the corresponding average overheads
given in \Cref{tab:related_work}.

\begin{figure*}
    \begin{minipage}[t]{\textwidth}
    \begin{table}[H]
        \begin{center}
            \caption{Snappy: Detailed table of Execution Times measurements  (\Cref{fig:perf1_abomonation})}
                \label{tbl:abomonation}
                    \begin{tabular}{|r|r|r|r|r|r|r|r|r|}
                        \hline 
                        \multicolumn{9}{|c|} {Abomonation V1 }\\ \hline  
                        Input     & \multicolumn{8}{|c|}{Execution Times [$\mu{}s$] (compress)}  \\
                        \cline{2-9}
                        bytes     & \SDRoB-FFI  &  $\sigma$  &\SDRoB-FFI &  $\sigma$   &\SDRoB-FFI &  $\sigma$   &\SDRoB-FFI&  $\sigma$    \\   
                                  &  opt-0   &              & opt-1    &             & opt-2   &             &opt-3   &             \\  \hline
                        $2^0$     & 2,36    & $\pm$21,38\% & 1,26  & $\pm$128,68\% & 1,09   & $\pm$369,58\% &1,09     & $\pm$55,19\%  \\\hline                                                                                                                
                        $2^3$     & 2,71    & $\pm$18,15\% & 1,30  & $\pm$29,44\%  & 1,08   & $\pm$37,72\%    &1,08   & $\pm$36,99\%    \\\hline                                                                                                                
                        $2^6$     & 5,06    & $\pm$11,92\% & 1,69  & $\pm$22,25\%  & 1,16   & $\pm$41,28\%    &1,15   & $\pm$37,77\%   \\ \hline                                                                                                                
                        $2^{9}$   & 21,72   & $\pm$5,87\%  & 4,83  & $\pm$17,84\%   & 1,38   & $\pm$34,57\%     &1,37 & $\pm$33,95\%    \\   \hline                                                                                                                
                        $2^{12}$  & 151,60  & $\pm$2,56\%  & 28,37  & $\pm$10,38\%   & 2,10   & $\pm$25,81\%     &2,10 & $\pm$23,71\%    \\   \hline                                                                                                                
                        $2^{15}$  & 1190,04 & $\pm$2,13\%  & 214,85  & $\pm$4,63\%    & 7,27   & $\pm$14,34\%  &7,32   & $\pm$14,34\%    \\    \hline                                                                                                                
                        $2^{18}$  & 9586,14 & $\pm$2,45\%  & 1770,13  & $\pm$2,99\%    & 104,06 & $\pm$7,61\%   &103,06 & $\pm$6,70\%   \\   \hline                                                                                                                
                        $2^{21}$  & 76327,93& $\pm$1,57\%  & 14688,60  & $\pm$2,41\%    & 1099,66& $\pm$2,61\%   &1100,45 & $\pm$2,49\%    \\   \hline
                                  & \multicolumn{8}{|c|}{Execution Times [$\mu{}s$] (uncompress)}  \\
                        \cline{2-9}
                        $2^0$     & 2,21        & $\pm$64,37\% & 1,20   & $\pm$122,08\%   & 1,01     & $\pm$281,91\% &1,00     & $\pm$39,60\% \\\hline                                                                                                                
                        $2^3$     & 2,52        & $\pm$18,29\% & 1,23   & $\pm$31,15\%     & 0,99     & $\pm$35,18\% &0,99     & $\pm$36,07\%  \\\hline                                                                                                                
                        $2^6$     & 4,73        & $\pm$10,93\% & 1,59   & $\pm$22,25\%     & 1,03     & $\pm$38,94\% &1,01     & $\pm$36,88\% \\ \hline                                                                                                                
                        $2^{9}$   & 21,17       & $\pm$5,91 \% & 4,60   & $\pm$18,21\%     & 1,11     & $\pm$33,51\% &1,07     & $\pm$36,54\%  \\   \hline                                                                                                                
                        $2^{12}$  & 150,54      & $\pm$2,66 \% & 27,74  & $\pm$8,76\%     & 1,49     & $\pm$31,92\% &1,39     & $\pm$32,79\%  \\   \hline                                                                                                                
                        $2^{15}$  & 1188,71     & $\pm$2,14 \% & 213,98 & $\pm$4,65\%     & 6,07     & $\pm$15,29\% &6,00     & $\pm$15,35\%  \\    \hline                                                                                                                
                        $2^{18}$  & 9574,67     & $\pm$2,46 \% & 1758,30    & $\pm$3,01   \% & 91,19    & $\pm$7,88\% &91,58     & $\pm$8,68\% \\   \hline                                                                                                                
                        $2^{21}$  & 76266,24    & $\pm$1,59 \% & 14561,62   & $\pm$2,43   \% & 1025,46  &$\pm$1,66\% & 1023,86    & $\pm$1,81\%  \\   
                        
                        \hline \hline 
                        \multicolumn{9}{|c|}{Abomonation V2} \\ \hline
                        Input     & \multicolumn{8}{|c|}{Execution Times [$\mu{}s$] (compress)}  \\
                        \cline{2-9}
                        bytes     & \SDRoB-FFI &  $\sigma$  &\SDRoB-FFI &  $\sigma$   &\SDRoB-FFI &  $\sigma$   &\SDRoB-FFI&  $\sigma$    \\   
                                  &  opt-0        &       & opt-1    &             &opt-2   &             &opt-3   &             \\  \hline
                        $2^0$     & 2,21     & $\pm$29,13\%& 1,14     & $\pm$24,88\%  & 1,16    & $\pm$343,80\% & 1,10  & $\pm$234,90\%  \\\hline                                                                                                                
                        $2^3$     & 2,19     & $\pm$14,92\%& 1,16     & $\pm$28,66\%  & 1,15    & $\pm$33,56\% & 1,08  & $\pm$32,85\%    \\\hline                                                                                                                
                        $2^6$     & 2,46     & $\pm$15,26\%& 1,23     & $\pm$23,15\%  & 1,21    & $\pm$33,60\% & 1,15  & $\pm$37,36\%  \\ \hline                                                                                                                
                        $2^{9}$   & 2,76     & $\pm$14,37\%& 1,44     & $\pm$20,05\%  & 1,40    & $\pm$31,28\% & 1,36  & $\pm$31,84\%    \\   \hline                                                                                                                
                        $2^{12}$  & 3,58     & $\pm$13,00\%& 2,24     & $\pm$19,37\%  & 2,12    & $\pm$25,50\% & 2,10  & $\pm$27,84\%    \\   \hline                                                                                                                
                        $2^{15}$  & 8,55     & $\pm$12,14\%& 7,54     & $\pm$12,71\%  & 7,36    & $\pm$18,49\% & 7,38  & $\pm$19,03\%    \\    \hline                                                                                                                
                        $2^{18}$  & 109,78   & $\pm$4,77\% & 104,18   & $\pm$6,51\%   & 105,98  & $\pm$8,49\% & 104,20& $\pm$6,73\%   \\   \hline                                                                                                                
                        $2^{21}$  & 1058,57  & $\pm$1,70\% & 1108,83  & $\pm$4,70\%  & 1112,27 & $\pm$2,07\% & 1107,01& $\pm$1,87\%   \\   \hline
                            & \multicolumn{8}{|c|}{Execution Times [$\mu{}s$] (uncompress)}  \\
                        \cline{2-9}
                        $2^0$     & 2,12     & $\pm$29,13\% & 1,07     & $\pm$102,90\% & 1,03      & $\pm$63,94\% & 1,00    & $\pm$78,57\% \\\hline                                                                                                                
                        $2^3$     & 2,10     & $\pm$14,92\% & 1,05     & $\pm$31,62\%  & 1,03      & $\pm$32,94\% & 0,99    & $\pm$46,59\%  \\\hline                                                                                                                
                        $2^6$     & 2,20     & $\pm$15,26\% & 1,06     & $\pm$21,61\%  & 1,03      & $\pm$43,21\% & 1,01    & $\pm$37,33\% \\ \hline                                                                                                                
                        $2^{9}$   & 2,29     & $\pm$14,37\% & 1,13     & $\pm$20,62\%  & 1,07      & $\pm$41,57\% & 1,07    & $\pm$40,21\%  \\   \hline                                                                                                                
                        $2^{12}$  & 2,67     & $\pm$13,00\% & 1,46     & $\pm$21,55\%  & 1,38      & $\pm$31,92\% & 1,40    & $\pm$33,27\%  \\   \hline                                                                                                                
                        $2^{15}$  & 7,12     & $\pm$12,14\% & 6,30     & $\pm$13,67\%  & 6,08      & $\pm$14,52\% & 6,15    & $\pm$22,39\%  \\    \hline                                                                                                                
                        $2^{18}$  & 93,67    & $\pm$4,77\%  & 92,20    & $\pm$6,96\%   & 94,11     & $\pm$8,28\%  & 90,25   & $\pm$7,93\% \\   \hline                                                                                                                
                        $2^{21}$  & 1004,03  & $\pm$1,70\%  & 1024,66  & $\pm$2,57\%   & 1022,25   & $\pm$1,71\%  & 1027,27 & $\pm$1,70\%  \\   \hline
                    \end{tabular}
        \end{center}
    \end{table}
    \end{minipage}
\end{figure*}

\begin{figure*}
    \begin{minipage}[t]{\textwidth}
    \begin{table}[H]
        \begin{center}
            \caption{Snappy: Detailed table of Execution Times measurements (\Cref{fig:perf1_serialization})}
                \label{tbl:snappy}
                \resizebox{\textwidth}{!}{
                    \begin{tabular}{|r|r|r|r|r|r|r|r|r|r|r|}
                        \hline 
                        Input     & \multicolumn{10}{|c|}{Execution Times [$\mu{}s$] (compress) }  \\
                        \cline{2-11}
                        bytes     & Baseline &  $\sigma$  &\SDRoB-FFI &  $\sigma$   &\SDRoB-FFI &  $\sigma$   &\SDRoB-FFI&  $\sigma$   & Sandcrust & $\sigma$   \\   
                                  &          &            & Abomonation &             &Bincode    &           &Bincode-2&                &         &            \\  \hline
                        $2^0$     & 0,11     & $\pm$102,43\% & 1,27            & $\pm$30,41\% & 1,24        & $\pm$16,82\% &0,99         & $\pm$26,65\% & 8,88    & $\pm$9,44\% \\\hline                                                                                                                
                        $2^3$     & 0,11     & $\pm$97,86\% & 1,23            & $\pm$27,37\% & 1,41        & $\pm$19,99\% &1,13         & $\pm$25,06\% & 8,78    & $\pm$9,17\%  \\\hline                                                                                                                
                        $2^6$     & 0,20     & $\pm$62,73\% & 1,30            & $\pm$21,69\% & 2,52        & $\pm$15,53\% &1,57         & $\pm$30,86\% & 8,92    & $\pm$7,67\% \\ \hline                                                                                                                
                        $2^{9}$   & 0,36     & $\pm$6,36\% & 1,55            & $\pm$10,99\% & 6,62        & $\pm$13,71\% &3,77         & $\pm$17,24\% & 9,44    & $\pm$3,18\%  \\   \hline                                                                                                                
                        $2^{12}$  & 0,84     & $\pm$22,29\% & 2,22            & $\pm$25,54\% & 35,06       & $\pm$6,97\% &19,89        & $\pm$9,03\% & 13,57   & $\pm$7,20\% \\   \hline                                                                                                                
                        $2^{15}$  & 2,41     & $\pm$20,36\% & 7,30            & $\pm$10,74\% & 249,71      & $\pm$4,86\% &148,37       & $\pm$4,76\% & 37,06   & $\pm$6,48\% \\    \hline                                                                                                                
                        $2^{18}$  & 20,00    & $\pm$30,98\% & 101,78          & $\pm$6,31\% & 1997,97     & $\pm$3,05\% &1222,46      & $\pm$3,43\% & 391,08  & $\pm$3,68\%\\   \hline                                                                                                                
                        $2^{21}$  & 246,23   & $\pm$4,31\% & 1075,22         & $\pm$2,43\% & 16318,68    & $\pm$2,42\% &10177,26     & $\pm$2,14\% & 2579,06 & $\pm$8,71\% \\   
                        \hline \hline
                            & \multicolumn{10}{|c|}{Execution Times [$\mu{}s$] (uncompress)}  \\
                        \cline{2-11}
                        $2^0$     & 0,09     & $\pm$85,08\% & 1,19    & $\pm$37,24\% & 1,21        & $\pm$50,42\% & 0,88     & $\pm$26,07\% & 8,86 & $\pm$8,02\% \\\hline                                                                                                                
                        $2^3$     & 0,08     & $\pm$105,91\% & 1,16    & $\pm$34,33\% & 1,46        & $\pm$16,82\% & 1,02     & $\pm$28,34\% & 8,78 & $\pm$9,58\%  \\\hline                                                                                                                
                        $2^6$     & 0,09     & $\pm$14,70\% & 1,17    & $\pm$19,37\% & 2,37        & $\pm$15,58\% & 1,39     & $\pm$30,36\% & 9,05 & $\pm$7,58\% \\ \hline                                                                                                                
                        $2^{9}$   & 0,10     & $\pm$15,53\% & 1,27    & $\pm$9,05\% & 6,38        & $\pm$13,84\% & 3,40     & $\pm$18,17\% & 11,28 & $\pm$2,64\%  \\   \hline                                                                                                                
                        $2^{12}$  & 0,19     & $\pm$70,87\% & 1,53    & $\pm$29,06\% & 34,66       & $\pm$7,02\% & 19,16    & $\pm$8,86\% & 29,77 & $\pm$4,60\%  \\   \hline                                                                                                                
                        $2^{15}$  & 1,08     & $\pm$30,27\% & 6,09    & $\pm$12,32\% & 248,15      & $\pm$4,88\% & 147,02   & $\pm$4,76\% & 115,99 & $\pm$3,33\%  \\    \hline                                                                                                                
                        $2^{18}$  & 17,39    & $\pm$45,23\% & 88,07   & $\pm$7,18\% & 1987,48     & $\pm$3,06\% & 1207,99  & $\pm$3,33\% & 905,96 & $\pm$3,80\% \\   \hline                                                                                                                
                        $2^{21}$  & 618,29   & $\pm$59,77\% & 993,67  & $\pm$1,72\% & 16257,90    & $\pm$2,42\% & 10166,90 & $\pm$2,17\% & 6478,21 & $\pm$6,17\%  \\   

                        \hline
                    \end{tabular}}
        \end{center}
    \end{table}
    \end{minipage}
\end{figure*}

\begin{figure*}[t]
    \begin{subfigure}[b]{0.5\textwidth}
      \centering
      \includegraphics[scale=0.500]{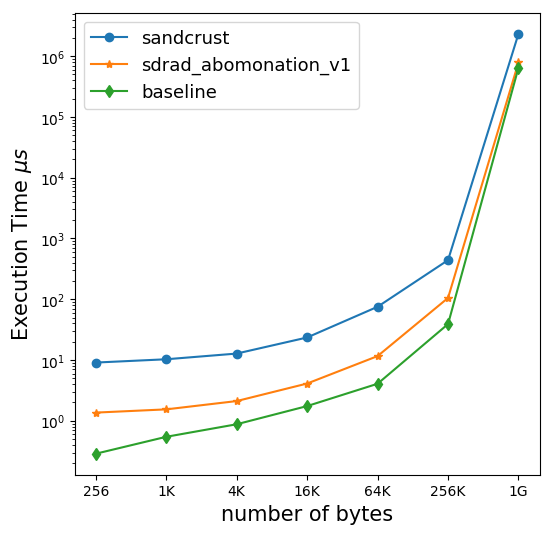}
      \caption{compress() benchmark}
      \label{fig:trust_compress}
    \end{subfigure}
    \hfill
    \begin{subfigure}[b]{0.5\textwidth}
      \centering
      \includegraphics[scale=0.500]{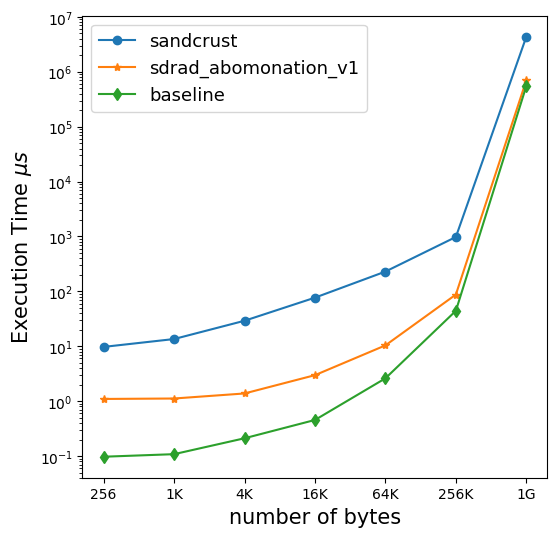}
      \caption{uncompress() benchmark}
      \label{fig:trust_decompress}
    \end{subfigure}
    \caption{Measuring snappy execution time for SDRaD-FFI}
    \label{fig:trust_compression}
  \end{figure*}

\begin{figure*}
    \begin{minipage}[t]{\textwidth}
    \begin{table}[H]
        \begin{center}
            \caption{Snappy: Detailed table of Execution Times measurements (\Cref{fig:trust_compression})}
                \label{tbl:trust}
                    \begin{tabular}{|r|r|r|r|r|r|r|}
                        \hline 
                        Input     & \multicolumn{6}{|c|}{Execution Times [$\mu{}s$] (compress)}   \\   
                        \cline{2-7}
                        bytes     & Baseline    &  $\sigma$     &\SDRoB-FFI   &  $\sigma$   &Sandcrust        &  $\sigma$          \\
                        $256$    & 0,29         & $\pm$56,55\%   & 1,36        & $\pm$24,47\% & 9,08       & $\pm$4,73\%         \\                                                                                                               
                        $1K$     & 0,54         & $\pm$11,11\%   & 1,54        & $\pm$2,49\% & 10,29      & $\pm$16,71\%          \\\hline                                                                                                                
                        $4K$     & 0,88         & $\pm$17,26\%   & 2,11        & $\pm$3,95\% & 12,77      & $\pm$7,95\%          \\\hline                                                                                                                
                        $16K$    & 1,75         & $\pm$15,39\%   & 4,10        & $\pm$7,58\% & 23,49      & $\pm$7,23\%         \\   \hline                                                                                                                
                        $64K$    & 4,07         & $\pm$60,39\%   & 11,61       & $\pm$4,26\% & 75,22      & $\pm$8,26\%         \\   \hline                                                                                                                
                        $256K$   & 39,07        & $\pm$70,67\%   & 105,19      & $\pm$2,97\% & 440,61     & $\pm$3,85\%          \\   \hline                                                                                                                
                        $1GB$    & 646610,97    & $\pm$1,07\%    & 791444,04   & $\pm$0,28\% & 2343068,37 & $\pm$0,65\%          \\   \hline   
                       $Geo Mean$& 11,37        &             & 29,08     &            &   155,28               &              \\\hline 
                       
                                    &             \multicolumn{6}{|c|}{Execution Times [$\mu{}s$] (uncompress)}                                 \\             
                        \cline{2-7}
                                    & Baseline   &  $\sigma$       &\SDRoB-FFI   &  $\sigma$       &Sandcrust     &  $\sigma$         \\   
                        $256$       & 0,10        & $\pm$7,40\%  & 1,10        & $\pm$47,15\%  & 9,74         & $\pm$1,66\%   \\\hline
                        $1K$        & 0,11        & $\pm$4,95\%  & 1,12        & $\pm$25,85\%  & 13,54        & $\pm$3,71\%   \\\hline
                        $4K$        & 0,21        & $\pm$9,50\%  & 1,38        & $\pm$4,28\%  & 29,16        & $\pm$12,67\%   \\\hline
                        $16K$       & 0,45        & $\pm$6,42\%  & 2,98        & $\pm$12,65\% & 76,88        & $\pm$25,37\%   \\\hline
                        $64K$       & 2,61        & $\pm$78,54\% & 10,38       & $\pm$4,75\%  & 229,61       & $\pm$4,15\%   \\\hline
                        $256K$      & 44,29       & $\pm$82,15\% & 87,53       & $\pm$6,23\%  & 979,81       & $\pm$1,25\%   \\\hline
                        $1GB$       & 550920,53   & $\pm$1,26\%  & 701423,77   & $\pm$0,34\%  & 4362162,32   & $\pm$0,68\%   \\\hline
                        $Geo Mean$   &  4,86               &             &     22,80            &              &   312,35           &  \\ \hline
                    \end{tabular}
        \end{center}
    \end{table}
    \end{minipage}
\end{figure*}



\end{document}